# *AI and Remote Sensing for Resilient and Sustainable Built Environments: A Review of Current Methods, Open Data and Future Directions*

Ubada El Joulani
Department of Electronic and Electrical Engineering
Brunel University of London
Ubada.Eljoulani@brunel.ac.uk

Tatiana Kalganova
Department of Electronic and Electrical Engineering
Brunel University of London
Tatiana.Kalganova@brunel.ac.uk

Stergios-Aristoteles Mitoulis
The Bartlett School of Sustainable Construction,
University College London
S.Mitoulis@ucl.ac.uk

Sotirios Argyroudis
Department of Civil and Environmental Engineering
Brunel University of London
Sotirios.Argyroudis@brunel.ac.uk

*Abstract* - **Critical infrastructure, such as transport networks, underpins economic growth by enabling mobility and trade. However, ageing assets, climate change impacts (e.g., extreme weather, rising sea levels), and hybrid threats ranging from natural disasters to cyber-attacks and conflicts pose growing risks to their resilience and functionality. This review paper explores how emerging digital technologies, specifically Artificial Intelligence (AI), can enhance damage assessment and monitoring of transport infrastructure. A systematic literature review examines existing AI models and datasets for assessing damage in roads, bridges, and other critical infrastructure impacted by natural disasters. Special focus is given to the unique challenges and opportunities associated with bridge damage detection due to their structural complexity and critical role in connectivity. The integration of SAR (Synthetic Aperture Radar) data with AI models is also discussed, with the review revealing a critical research gap: a scarcity of studies applying AI models to SAR data for comprehensive bridge damage assessment. Therefore, this review aims to identify the research gaps and provide foundations for AI-driven solutions for assessing and monitoring critical transport infrastructures.**

## 1. Introduction

Transport networks are crucial for the integrity of the economy and social health of any region in the world, thus maintaining them in good condition is of high importance. Climate change is having a big impact on transport networks as well, as common climate threats include large precipitations, high temperatures and rising sea levels, which lead then to biophysical impacts such as floodings, erosion, and urban heat islands, which reduces road safety and durability (de Abreu et al., 2022). There are also direct impacts that refers to the actual damage to the infrastructures and indirect damages due to the cascading events (Rebally et al., 2021). Consequently, fostering climate-resilient infrastructure is becoming essential for the economic prosperity and social coherence of any country (Argyroudis et al., 2022), aligning with the United Nations Sustainable Development Goals (SDGs) (United Nations, 2015).

Given these threats, critical infrastructures require quick damage assessment to enable informed decision making and on time restoration avoiding cascading impacts. This need is especially highlighted in challenging zones, such as areas under war or other disruptive events. The use of remote sensing technologies and satellites is crucial here, as data collection in these areas is often defined by security risks and restricted access making on-ground data hard to obtain (Zhao & Morikawa, 2024).

Key methods to address this challenge involve damage assessment using satellite images, which can be sourced from open-access platforms or commercial providers. A prominent example is the ESA (European Space Agency) Sentinel mission, which provides valuable data through radar imaging (Sentinel-1) and multispectral high-resolution imaging (Sentinel-2) (*European Space Agency*, 2025). For instance, researchers have developed multi-scale approaches that integrate Sentinel-1 SAR images with high-resolution imagery and deep learning for rapid post-disaster infrastructure damage detection (Kopiika et al., 2025).

While there have been previous studies on damage detection, they have often focused on single transport infrastructures such as roads, bridges (Santaniello & Russo, 2023) or buildings in isolation. Existing literature reviews have also covered related topics, for example, (Abedi et al., 2023) provided a systematic review of Machine Learning for general civil infrastructure damage using methods like vibration and image analysis, while (Abduljabbar et al., 2019) presented a broader overview of AI applications across the transport sector without a specific focus on structural damage.

However, a comprehensive review is needed to compare current AI models and datasets for assessing road and bridge damage, as this is missing from the literature. Crucially, this review is motivated by another identified gap in the literature: while satellite technology like Synthetic Aperture Radar (SAR, a satellite-radar imaging technique that uses motion of a radar antenna to create high resolution images of the earth surface) is used for monitoring, its integration with advanced AI models for holistic bridge damage assessment remains largely unexplored. This review aims to address this gap. It will synthesize the latest emerging technologies and AI models, from the detection of localized road potholes to wider regional damage assessments, providing a foundation for developing AI-driven solutions that enhance the monitoring and resilience of critical transport infrastructures.

The rapid adoption of these technologies necessitates a careful consideration of ethical AI principles (Díaz-Rodríguez et al., 2023) (Radanliev, 2025). These concerns include fairness, transparency, privacy, and accountability. AI models for damage assessment could cause societal inequalities if trained on biased datasets. An AI system trained predominantly on urban or affluent area imagery might underperform in rural regions, leading to inequitable allocation of repair resources and marginalization of vulnerable groups. This raises the question for accountability, which demands mechanism to ensure responsibility for an AI system's outcome and provide compensation when its decision cause harm. Furthermore, the system must be transparent and explainable, making their functionality clear and understandable to build and maintain user trust. The use of high-resolution satellite imagery also introduces significant privacy and data governance concerns that must be addressed to protect individuals and ensure data is used responsibly.

The increasing use of AI in managing critical infrastructure demands significant policy and risk management reform, as some current regulations are inadequate, where rather than a single, universally accepted legal framework, countries are adopting varied approaches. The European Union has passed a comprehensive EU AI Act (*Artificial Intelligence Act: MEPs Adopt Landmark Law | News | European Parliament*, 2024), while the UK is pursuing a more flexible and principles-based approach (*A Pro-Innovation Approach to AI Regulation*, 2023). There is need for new policies that standardize data quality, model validation and operational transparency, as poor data quality leads to flawed and unreliable AI models (*What Is Data Governance? | IBM*, 2025). In terms of transparency, there is a global push to make AI systems more transparent and explainable so that they can be trusted and held accountable(*What Is Explainable AI (XAI)? | IBM*,

2025). It is also crucial to address the emerging landscape of AI-generated threats. The same technology can be used for malicious ends, and citing few examples, this includes Bioterrorism, Unleashing AI agents, persuasive AIs and Concentration of Power (Hendrycks et al., 2023) and other found in this publication (Janjeva et al., 2024). Therefore, the successful application of these technologies for infrastructure assessment requires navigating the challenges of ensuring ethical performance, establishing robust governance and policy, and safeguarding the assessment process from digital interference.

## 2. Methodology

This review paper employs a systematic approach to evaluate existing research and compare the different findings and applications of AI models and datasets availability. While the literature demonstrates significant literature, particularly in road damage detection, our initial analysis confirmed a scarcity of research combining AI, SAR, and bridge damage assessment. Therefore, this review aims to provide a comprehensive evaluation of current findings and highlight directions for future research.

Articles were included based on a series of criteria which includes the relevance of AI models and their application for damage assessment on transport infrastructures (roads, bridges, etc.), availability of the datasets that correspond to the tables' columns (i.e. for AI models, the accuracy). As for the eligibility criteria for article searching, we considered the most recent articles, including articles up to 10 years old, except for same cases where articles were scarce. The language of these articles is English, for ease of comparison and readability. Both Scopus and Google Scholar were utilised for the identification and review of relevant articles. The search terms are the titles identified for each table corresponding to a specific research question. We excluded articles that are not relevant to the research question identified for each table.

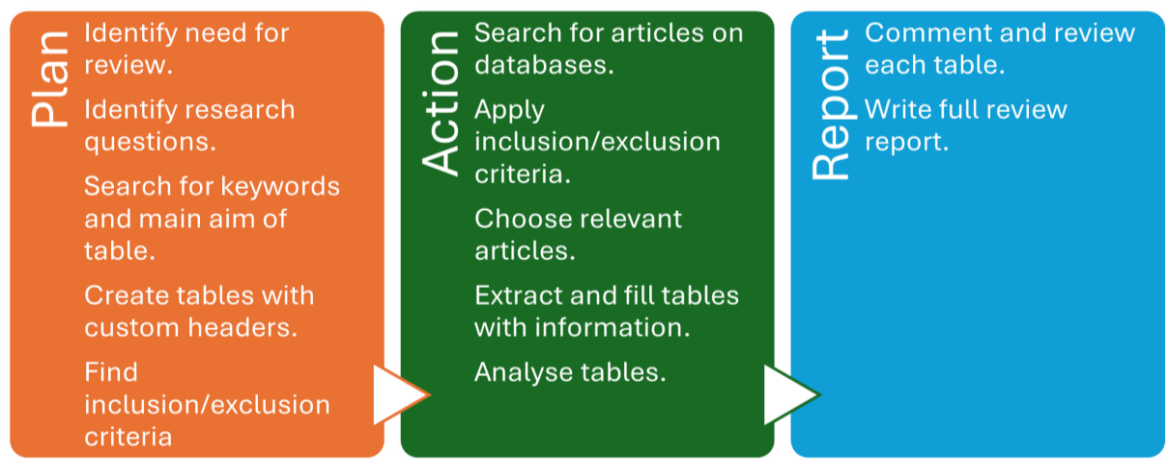

*Figure 1. Review process diagram methodology*

For wider damage assessment, we looked into technologies that use satellite imagery as well, and more specifically SAR (Kopiika et al., 2025) (Nettis et al., 2023) (Markogiannaki et al., 2022). This technology has been used with varied methodology depending on applications, such as MTInSAR (Multi-Temporal Interferometric SAR), InSAR (Interferometric SAR) and D-TomoSAR (Differential Tomographic SAR). For instance, MTInSAR has been used for monitoring of structural deformation in bridge portfolios, as demonstrated by (Nettis et al., 2023). InSAR has also been used for similar applications, such as long-term deflection and thermal dilation of bridges (Jung et al., 2019). D-TomoSAR is similar to the previously mentioned methods, but it uses multiple radar images acquired from different viewing angles to create 3D model of the deformed infrastructure. In (Markogiannaki et al., 2022), the authors have used D-TomoSAR for monitoring landmark bridge using displacement products deformation trends. Another application of SAR includes using coherence products (correlation of radar signal between two or more acquisitions of the same area) for assessing the damage on infrastructures, as demonstrated by (Kopiika et al., 2025) and (Sun et al., 2020), where Coherence Change Detection (CCD) has been used, in which two temporal high-resolution SAR images are compared to detect and measure changes to a specific geographic area, as described in Figure 2.

*Table 1. SAR RGB decomposition (Schultz, 2021)*

| Colour | Band | Polarization | Small contribution to pixel indicates | Large contribution to pixel indicates |
|---|---|---|---|---|
| Red | Co-Pol (VV) | Surface scattering (polarized/simple) | Smooth surface | Rough surface |
| Green | Cross-Pol (VH) | Volume scattering (depolarized/random) | Low volume (water, roads, plowed or newly planted fields) | High volume (trees, buildings, mature crops, built up areas) |
| Blue | C-Pol when Cross-Pol near 0dB | Surface scattering when volume scattering is very low | Scattering measurable in red channel, no value | Co-Pol backscatter values near -24dB (smooth water, roads) |

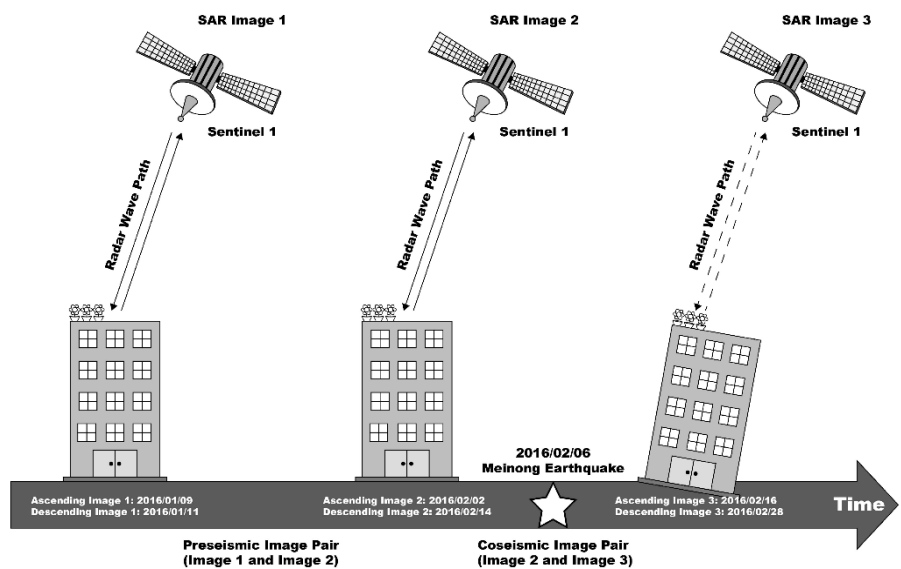

*Figure 2. SAR acquisition for CCD (Lu et al., 2018)*

Some applications of SAR for have used RGB composite images using various SAR data parameters, where different polarizations and frequencies of radar signals are combined into a multi-band visualization (Heiselberg, 2020). These RGB composite images can be used in AI models using computer vision algorithms, improving the capabilities of the system to assess damage for transport infrastructures. In Table 1, the difference between colours is shown, with what features can be identified in each channel.

## 3. Results and Discussion

In this section, an analysis of current available AI models and datasets in this field are carried out along with an investigation on robust solutions for data collection. Through the literature review, the emerging digital technologies and system resilience will be also explored. The findings are presented in table style, along with a discussion regarding the damages on civil infrastructures that have been studied, along with available datasets and AI models deployed. Consequently, an analysis on the use of SAR for damage detection is presented, and lastly a discussion on natural hazards related to damaged civil infrastructures.

### 3.1 Damages in critical civil infrastructure used for analysis in AI approaches

Table 2 summarises the most common types of civil infrastructure damage that have been identified, which allows us to have a clear view of what the most common ones are. Although some have used different criteria, such as good, fair and poor (referring to the state of the road) as performed by (Ma et al., 2017). When searching for publications, we found that most of the damage detection models were applied on road infrastructure instead of buildings. When searching on Google Scholar "Road Damage Detection" since 2012 to 2025 the search gives us more papers compared to when searching "Building Damage Detection". This explains that there has been more research and applications on road infrastructures, detecting cracks and potholes, compared to buildings. The above table tells us that most of the cracks identified are classified into lateral, longitudinal, alligator, and other general cracks.

*Table 2. Damages in critical civil infrastructure used for analysis in AI approaches*

| Author / Year | Roads | | | | |
|---|---|---|---|---|---|
| | **Longitudinal cracks** | **Lateral cracks** | **Alligator cracks** | **Cracks** | **Other road components** |
| (Paramasivam et al., 2024) | | | | x | Potholes |
| (Zeng & Zhong, 2024) | x | x | | x | Potholes, mesh cracks |
| (Guo & Zhang, 2022) | x | x | | x | Mesh cracks, pothole, longitudinal and lateral construction joint, crosswalk blur, white line blur. |
| (Stricker et al., 2021) | x | x | x | x | Patches, scratches, bleeding, manholes, curb, cobblestone, drill holes, vegetation, joints, water drains. |
| (Du et al., 2021) | x | x | x | x | Patches, nets, manholes |
| (Mei & Gül, 2020) | | | | x | |
| (Majidifard et al., 2020) | x | x | x | x | Reflective crack, block crack, sealed reflective crack, lane long. crack, sealed long. crack |
| (Hegde et al., 2020) | x | x | x | | |

| | | | | |
|---|---|---|---|---|
| (Angulo et al., 2019) | x | | x | x |
| (Stricker et al., 2019) | | | x | x |
| (Weng et al., 2019) | x | x | x | x |
| (Maeda et al., 2018) | x | x | x | |
| (Dorafshan et al., 2018) | | | | x |
| (Ma et al., 2017) | | | | | Good, Fair, Poor |
| (Ouma & Hahn, 2017) | | | x | |
| (L. Zhang et al., 2016) | | | | x |
| (Shi et al., 2016) | | | | x |
| (L. Li et al., 2014) | x | x | x | |
| (Oliveira & Correia, 2014) | x | x | | x |
| (Zou et al., 2012) | | | | x |

**Buildings**

| | Minor | Major | Destroyed | Ruin | Other |
|---|---|---|---|---|---|
| (Bhardwaj et al., 2025) | | | | | Damage, No damage |
| (Braik & Koliou, 2024) | x | x | x | | Undamaged |
| (V.V. et al., 2024) | | | | | Damaged, undamaged |
| (Kaur et al., 2023) | x | x | x | | Undamaged |
| (Y. Zhang et al., 2023) | | | x | x | |
| (C. Liu et al., 2022) | | | x | | Debris, Spalling, Cracking |
| (C. Wu et al., 2021) | x | x | x | | |
| (Y. Wang et al., 2022) | x | x | x | | |
| (Weber & Kané, 2020) | x | x | x | | |
| (Gupta et al., 2019) | x | x | x | | |
| (J. Z. Xu et al., 2019) | | | | | UNOSAT 5-level scale |

Regarding damage criteria for buildings, we can observe from Table 2 that, comparably for the roads, it is common to classify the following: "Minor", "Major" and "Destroyed". Upon analysis of these papers, it was revealed that these classifications are used due to the widespread use of the "xBD" dataset, which is a large-scale dataset of building damage assessment used for humanitarian relief and disaster rescue.

## 3.2 AI approaches used for damage detection on infrastructures

A wide range of AI models have been applied in damage detection algorithms on civil and transport infrastructures. These models include both traditional machine learning methods such as Random Forest (Shi et al., 2016), as well as more advanced deep learning models, for instance CNNs (Convolutional Neural Network) (Waseem Khan et al., 2025), (Paramasivam et al., 2024), (Majidifard et al., 2020).

In Table 3, a series of these AI models are presented and compared. The references are presented in the first column, the name of the model in the second column, and lastly the source of the data, commonly Terrestrial or Satellite. Here the infrastructures are roads, buildings, bridges, and other (which is only a steel structural model, to show an example of simulation of damage detection).

Table 3. Overview of AI approaches for damage detection in buildings, bridges and roads, including algorithms, datasets, and data sources. Abbreviations[1]

| Author / Year | Algorithm | Dataset | Source |
|---|---|---|---|
| **Roads** | | | |
| (Waseem Khan et al., 2025) | YOLOv9s-Fusion | RDD2022 | Terrestrial |
| (Shakhovska et al., 2024) | YOLO_tinyv4 | Potholes or Cracks on Road Image Dataset | Terrestrial |
| (Zanevych et al., 2024) | YOLOv11+FPN+Crad-CAM | Multiple publicly available | Terrestrial |
| (Khan et al., 2024) | Faster R-CNN, YOLOv5, SSD MobileNet V1, EfficientDet D1 | RDD2022 | Terrestrial |
| (Ji et al., 2024) | LRDD-YOLO | Pothole dataset, Road Damage Dataset | Terrestrial |
| (Paramasivam et al., 2024) | Faster R-CNN | Custom | Terrestrial |
| (Y.; Li et al., 2024) | RDD-YOLO | RDD2022 | Terrestrial |
| (Zeng & Zhong, 2024) | YOLOv8-PD | RDD2022 | Terrestrial |
| (J. Chen et al., 2024) | LAG-YOLO | RDD2020 | Terrestrial |
| (Ni et al., 2023) | YOLOv7 | RDD2022 | Terrestrial |
| (Guo & Zhang, 2022) | YOLOv5s | RDD2020 | Terrestrial |
| (Arya, Maeda, Ghosh, Toshniwal, Mraz, et al., 2021) | YOLOv5 | RDD2020 | Terrestrial |
| (Du et al., 2021) | YOLOv3 | LIST dataset | Terrestrial |
| (Hegde et al., 2020) | u-YOLO with EM&EP | GRDDC | Terrestrial |
| (F. Yang et al., 2020) | FPHBN | CRACK500, GAPs384, Cracktree200, CFD, (Aigle-RN & ESAR & LCMS) | Terrestrial |
| (Mei & Gül, 2020) | ConnCrack (GANs) | EdmCrack600 | Terrestrial |
| (Majidifard et al., 2020) | YOLOv2, Faster RCNN | PID (pavement image dataset) | Terrestrial |
| (Angulo et al., 2019) | RetinaNet | Custom | Terrestrial |
| (Weng et al., 2019) | Edge detector and segmentation | Custom | Terrestrial |
| (Stricker et al., 2019) | ResNet34 160x160 | GAPs V2 | Terrestrial |
| (Dorafshan et al., 2018) | AlexNet DCNN | SDNET2018 | Terrestrial |
| (Maeda et al., 2018) | SSD-Inception v2 | RoadDamageDetector | Terrestrial |

---

[1] EM: Ensemble Model, EP: Ensemble Prediction, ASPP: Atrous Spatial Pyramid Pooling, U-BDD++: Improved unsupervised building damage detection, FV: Fisher vector, FPN: Feature Pyramid Network, RSF: Random Structured Forests, RDF: Random Decision Forests, BPNN: Back-Propagation NN, FPHBN: feature pyramid and hierarchical boosting network, SCWT: Synchrosqueezing Continuous Wavelet Transform, YOLO: You Only Look Once, CNN: Convolutional Neural Network, GAN: Generative Adversarial Network, SSD: Single Shot Detector

| (Ouma & Hahn, 2017) | Fuzzy c-means | Custom | Terrestrial |
| --- | --- | --- | --- |
| (Ma et al., 2017) | FV-CNN | Custom – Google street view | Terrestrial |
| (Shi et al., 2016) | CrackForest (RSF+RDF) | CFD, AigleRN | Terrestrial |
| (L. Zhang et al., 2016) | Convnets | Custom | Terrestrial |
| (L. Li et al., 2014) | BPNN | ARAN dataset | Terrestrial |
| **Buildings** | | | |
| (Bhardwaj et al., 2025) | ResNet, U-Net | Custom | Satellite |
| (Y. Yang et al., 2024) | RNN | Custom | Satellite |
| (Braik & Koliou, 2024) | CNN | xBD | Satellite |
| (C. Wang et al., 2024) | Bayesian networks | xBD | Satellite |
| (V.V. et al., 2024) | U-Net | Custom | Satellite |
| (Kaur et al., 2023) | Hierarchical Transformer | xBD, Ida-BD, LEVIR-CD | Satellite |
| (Y. Zhang et al., 2023) | U-BDD++ | xBD | Satellite |
| (Y. Wang et al., 2022) | DNN | xBD | Satellite |
| (C. Liu et al., 2022) | LA-YOLOv5 | GDBDA | Terrestrial |
| (Weber & Kané, 2020) | Mask R-CNN with FPN | xBD | Satellite |
| (Gupta et al., 2019) | ResNet50, CNN | xBD | Satellite |
| (J. Z. Xu et al., 2019) | AlexNet CNN | Custom | Satellite |
| **Bridges** | | | |
| (Abubakr et al., 2024) | Xception Vanilla | CODEBRIM | Terrestrial |
| (Santaniello & Russo, 2023) | SCWT & ResNet with signal splitting | Z24 bridge | Terrestrial |
| (Gao et al., 2023) | GoogleNet | Crack-detection | Terrestrial |
| (Ni et al., 2023) | YOLOv7 | RDD2022 | |
| (Tazarv et al., 2022) | Mask R-CNN | RC-bridge | Terrestrial |
| (Mundt et al., 2019) | MetaQNN and ENAS | CODEBRIM | Terrestrial |
| (H. Xu et al., 2019) | CNN with ASPP | Crack-detection | Terrestrial |

In the case of buildings, most of these publications are related to building damage related to natural disasters, as shown in (Y. Yang et al., 2024) for earthquake and (Bhardwaj et al., 2025), (C. Wang et al., 2024), (Braik & Koliou, 2024) and (Kaur et al., 2023) for hurricanes. These recent publications relied mainly on the publicly available dataset xBD (Gupta et al., 2019). (Y. Zhang et al., 2023) have presented an innovative model, where the authors have achieved an F1 score of 0.582 for the tasks of localization and segmentation, and an F1 score of 0.638 for the tasks of damage classification. Here the data used consisted of unlabelled pre and post disaster satellite images pairs. Using satellite is not always the optimal solution due to complexity, so the authors have implemented a novel self-supervised framework, named U-BDD++. Other findings, (C. Liu et al., 2022), show higher accuracy on a different dataset, such as the GDBDA(Ground-level Detection in Building Damage Assessment), where an average (between different classes) F1 score of 0.911 was achieved, using a improved version of YOLOv5 object detection model. A term has been found for the application of Artificial Intelligence to geospatial data from remote sensors such as satellites, aerial drones, and this is GeoAI (Agbaje et al., 2024). GeoAI brings a big potential for Rapid and scaled-up building damage assessment.

As for bridges, we observe that most of the publications used CNNs, deep learning models, for damage identification. Some authors have used an improved version of Convolutional Neural Networks, such as Xception, a spatial architecture that is more powerful with less over-fitting problems than current popular CNNs (Abubakr et al., 2024). The authors have utilized Xception model and Vanilla model, achieving respectively an accuracy of 0.9495 and 0.8571 for defect classification of concrete bridges. Other authors have experimented with different models, such as Meta-QNN (Mundt et al., 2019), a meta-modelling algorithm based on reinforcement learning that generated higher performance architectures automatically, and Synchrosqueezing Continuous Wavelet Transform with deep learning (Santaniello & Russo, 2023), using acceleration responses for multi-class damage detection.

When it comes to roads, there has been a lot of competitions, such as the Global Road Damage Detection, which happened on multiple occasions, like in 2020 and 2022. In fact, we have presented the relative datasets in the below tables, under RDD2020 and RDD2022. There has been some variation to these datasets and competitions, such as the Optimized Road Damage Detection Challenge ([ORDDC'2024](ORDDC'2024)) or the Crowdsensing-based Road Damage Detection Challenge (CRDDC) (Arya et al., 2022). From the table we understand that most of the models used are based on YOLO (You Only Look Once) models, which are two stage detectors (Redmon et al., 2016): in the first pass it generates the potential object locations, and in the second pass it refines these proposals. A recent study presents a model specifically developed for road damage detection, where the authors based on a previous object detection model YOLOv8n, have proposed an improved version, YOLOv8-PD for Pavement Distress, demonstrating lower computational load and higher detection accuracy (Zeng & Zhong, 2024). Most recent versions have also been used such as YOLOv11 (Zanevych et al., 2024) and YOLOv9 (Waseem Khan et al., 2025), and recently, as the weight of the models are being considered more and more, particularly for edge applications, lighter versions are also being considered, such as YOLO_tinyv4 (Shakhovska et al., 2024).

An experiment have been conducted on simulated structures, such as an eight-level steel frame structure, where in (Jiang et al., 2022), a two-stage structural damage detection method is used (a 1D-CNN model in the first stage to extract the damage features, and a SVM model to quantify the damage), and achieved a high accuracy of 0.9988. However, it has not yet been applied to real world infrastructure, where additional factors influence the performance. Lastly, the majority of these papers have relied on terrestrial data, with limited use of satellite imagery, despite its value in scenarios where access to transport infrastructure is restricted.

### 3.3 Datasets used for infrastructure damage detection

In the previous sections, the AI models have been presented, along with what datasets have been used. In this part, these datasets are more deeply analysed. In Table 4 the datasets for the different infrastructures are presented. We can observe how the section for roads is bigger compared to buildings and bridges. This is because the datasets for roads are easier to create compared to buildings and bridges, which requires more sophisticated and advanced acquisition techniques, as demonstrated later in the table about technologies used for data collection. We can observe that to create a road dataset, a smartphone with a camera is sufficient. Furthermore, there have been numerous competitions for road damage detection like the RDD2020 and RDD2022, which had a huge success and motivated for more advanced datasets, i.e. including other countries' roads to improve the model. For instance, in RDD2020 dataset (Arya, Maeda, Ghosh, Toshniwal, & Sekimoto, 2021), the data was collected from three different countries: India, Japan and Czech Republic. However, in RDD2022 dataset (Arya et al., 2024), the data was collected from six countries: India, Japan, Czech Republic, Norway, the United States and China, with more than 55,000 instances of road damage.

*Table 4. Datasets of damaged infrastructures used for detection.*

| Authors/Year | Dataset | Classes | No. of Images | Images resolution |
|---|---|---|---|---|
| **ROADS** | | | | |
| (Shakhovska et al., 2024) | Potholes or Cracks on Road Image Dataset | Longitudinal, transverse, alligator crack, potholes, rutting, surface distress. | 1,000+ | 1920x1080 |
| (Arya et al., 2024) | RDD2022 CRDDC2022 | Longitudinal, Transverse, Alligator cracks, Potholes. | 47,420 | 512x512, 600x600, 720x720, 3,650x2044 |
| (Du et al., 2021) | LIST | Crack, Pothole, Net, Patch-Crack, Patch-Pothole, Patch-Net, Manhole. | 45,788 | 1,920x1080 |
| (Arya, Maeda, Ghosh, Toshniwal, & Sekimoto, 2021) | RDD2020 | Longitudinal cracks, Transverse cracks, Alligator cracks, and Potholes. | 26,336 | 600x600, 720x720 |
| (Stricker et al., 2021) | GAPs 10m | 22 classes[2]. | 394 | 5,030x11,505 |
| (F. Yang et al., 2020) | Crack500 | Crack. | 500 | 2,000x1500 |
| (Mei & Gül, 2020) | EdmCrack600 | Crack. | 600 | 1,920x1080 |
| (Majidifard et al., 2020) | PID | Block, Lane longitudinal, Longitudinal, Sealed Longitudinal, Pothole, Alligator, Sealed reflective, Reflective, Transverse. | 7,237 | 640x640 |
| (Stricker et al., 2019) | GAPs v2 | Intact, Cracks, Applied patches, Inlaid patches, Potholes, Open joints. | 2,468 | 1,920x1080 |
| (Angulo et al., 2019) | Modified RDD2018 | Wheel mark, Construction joint long., Equal interval, Construction joint lat., Partial/Overall pavement, Bump/Rutting, Crosswalk blur, White line blur. | 18,034 | 600x600 |
| (Weng et al., 2019) | G45 | Transverse, Longitudinal, Block, Alligator | 217 | 2,048x1,536 |
| (Dorafshan et al., 2018) | SDNET2018 | Cracked, Non-cracked | 56,000 | 256x256 |
| (Ma et al., 2017) | NYCDT | Poor, Fair, Good. | 711,520 | 640x640 |
| (Ouma & Hahn, 2017) | Custom | Illumination and light intensity variations, Background asphalt variations, Cracks, Oil stains, Patches, Pebbles, Shadows, other. | 75 | 1,080x1,920 |
| (Shi et al., 2016) | CFD | Crack, Non-crack. | 118 | 480x320 |
| (L. Zhang et al., 2016) | Custom | Crack, Non-crack. | 500 | 3,264x2,448 |
| (L. Li et al., 2014) | Custom | Alligator crack, Linear crack:(Longitudinal, Transversal crack). | 400 | n/a |
| (Oliveira & Correia, 2014) | CrackIT | Crack, Non-crack. | 84 | 1,536x2,048 |
| **BUILDINGS** | | | | |
| (Y. Yang et al., 2024) | Custom | Collapsed, Heavily damaged, Needs demolished, Slightly damaged | 13 | n/a |
| (C. Wang et al., 2024) | Custom | No Damage, Minor, Moderate, Severe, Destroyed | 2,472 | n/a |
| (V.V. et al., 2024) | Custom | Damaged, undamaged | 50 | 512x512 |
| (C. Liu et al., 2022) | GDBDA | Debris, Collapse, Spalling, Crack. | 8,340 | 800x800 |
| (Gupta et al., 2019) | xBD | No damage, Minor damage, Major damage, Destroyed, Unclassified. | 22,068 | 1,024x1,024 |
| (J. Z. Xu et al., 2019) | Custom | No damage, Possible Damage, Moderate Damage, Severe Damage, Destroyed | 75,468 | 0.3 GSD |
| **BRIDGES** | | | | |
| (Flotzinger et al., 2023) | Dacl10k | 12 classes[3]. | 9,920 | Min: 336x245 Max: 6,000x5,152 |
| (Santaniello & Russo, 2023) | Z24 | Undamaged, 20mm, 40mm, 80mm, 95mm displacement. | 1,422 | Time-series |

---

[2] Void, Inlaid patch, Applied patch, Scaled crack, Crack, Open joint, Pothole, Raveling, Scratch, Bleeding, Road marking, Surface water drain, Manhole, Expansion joint, Curb, Cobblestone, Drill hole, Object mobile, Object fixed, Joint, Road verge, Vegetation, Induction loop, Normal.
[3] Crack, Alligator crack, Efflorescence, Rockpocket, Washouts concrete corrosion, Hollowareas, Spalling, Restformwork., Wetspot, Rust, Graffity, Weathering, ExposedRebars, Bearing, Expansion joint, Drainage, Protective equipment, Joint tape.

| | | | | |
|---|---|---|---|---|
| (H. Xu et al., 2019) | Crack-detection | Crack, Non-crack. | 6,069 | 224x224 |
| (Mundt et al., 2019) | CODEBRIM | Crack, Spallation. Efflorescence, Exposed Bars, Corrosion. | 1,590 | 2,592x1,944 to 6,000x4,000 |
| (Dorafshan et al., 2018) | SDNET2018 | Cracked, Non-cracked. | 56,000 | 256x256 |

From the table we can observe the disparity between image sizes across datasets. Some images were collected using specific advanced systems with very high images resolution, such as "Mobile mapping system" named S.T.I.E.R. and RoadSTAR (Stricker et al., 2021), which have been used in Austria, Switzerland and Germany.

For buildings there are fewer datasets, but there is one dataset that used satellite images that is very extensive, including approximately 22 thousand images over 45 kilometres squared of polygon labelled pre and post disaster imagery, the xBD dataset (Gupta et al., 2019). The custom datasets are collected from the xBD dataset, for specific damages and specific locations, depending on the area of interest. Now there is a recent published one called "Bright" which has data about damaged buildings related to natural disasters (H. Chen et al., 2025).

In the context of bridges, there is a noticeable scarcity of publicly available image datasets specifically capturing overall structural damage. This scarcity is particularly acute for datasets suitable for advanced remote sensing techniques like SAR, which directly hinders the development and validation of corresponding AI models. However, several datasets focused on localised defects, particularly concrete cracks in bridge components, are available, such as the widely used CODEBRIM dataset (Mundt et al., 2019). Vibration based approaches have also been investigated for bridge damage assessment. For example, (Santaniello & Russo, 2023) applied deep neural networks to time-frequency representations of vibration signals to detect structural damage. Their study utilized the Z24 dataset, a well-known benchmark in the field; however, this dataset is not publicly accessible, limiting its broader use in comparative studies. Another notable dataset for bridge damage detection is DACL10 (Flotzinger et al., 2023), a comprehensive dataset comprising 9,920 images collected from real-world bridge inspections. It supports multi-label semantic segmentation and includes annotations for 12 damage types across 6 distinct bridge components, making it a valuable resource for developing and evaluating deep learning models in realistic inspection scenarios.

We iterate here the importance of monitoring these structures, like bridges and roads, and identifying the right dataset and model is crucial for efficient restoration works, traffic load management and avoiding disruptions on major routes.

Table 5 shows some samples of the data/images in the different roads datasets here showed in Table 4. The images showed are taken randomly from different classes. In the GAPs 10m dataset by (Stricker et al., 2021), a system of high-resolution imaging was used, and we can see the sample images in Table 5. Another example is the building dataset xBD (Gupta et al., 2019), which by looking at the table of images, we can understand that the authors have used multi-band satellite imagery. In summary, this table shows some samples of how the data looks like, without the need of searching the dataset and looking at the images. In Table 6, samples of the bridge datasets used for damage detection are presented as well.

*Table 5. Samples images from road datasets and aerial/satellite*

| Author/Year | Open-source Dataset Name | Samples |
|---|---|---|
| (Arya et al., 2024) | RDD2022 | |
| (Du et al., 2021) | LIST | |
| (Arya, Maeda, Ghosh, Toshniwal, & Sekimoto, 2021) | RDD2020 | |
| (Stricker et al., 2021) | GAPs 10m | |
| (F. Yang et al., 2020) | Crack500 | |
| (Mei & Gül, 2020) | EdmCrack600 | |
| (Gupta et al., 2019) | xBD | |
| (Stricker et al., 2019) | GAPs v2 | |
| (Angulo et al., 2019) | Modified RDD2018 | |
| (Dorafshan et al., 2018) | SDNET2018 | |

| (Shi et al., 2016) | CFD | 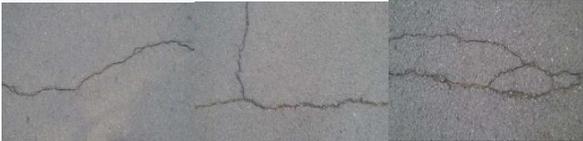 |

Table 6. Samples of images from bridge datasets

| Author/Year | Classes | No. of images | Samples |
|---|---|---|---|
| (IADF TC & GRSS IEEE, 2025) DOTA | Multiple classes, including Bridge | 11,268 | 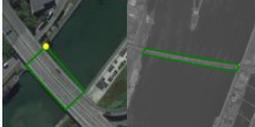 |
| (IADF TC & GRSS IEEE, 2025) Bridge Dataset | Bridges | 500 | 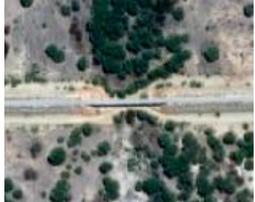 |
| (IADF TC & GRSS IEEE, 2025) AID | Multiple classes, including Bridge | 10,000 | 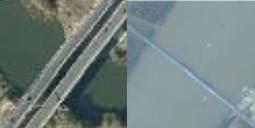 |
| (Flotzinger et al., 2023) Dacl10k | 12 classes (see footnote 2 above) | 9,920 | 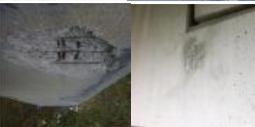 |
| (H. Xu et al., 2019) Crack-detection | Crack, Non-crack | 6,069 | 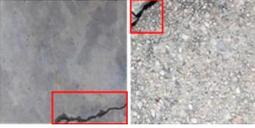 |
| (Mundt et al., 2019) CODEBRIM | Crack, Spallation, Efflorescence, Exposed Bars, Corrosion | 1,590 | 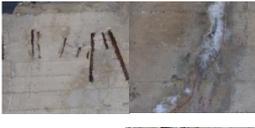 |
| (Dorafshan et al., 2018) SDNET2018 | Cracked, Non-cracked | 230 | 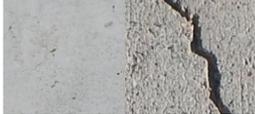 |

We looked at what datasets about general transport infrastructures are available previously in Table 4 however here in Table 6 we are visualizing sample images of the damaged bridges datasets we previously saw. As shown in the table, most of these datasets concern concrete cracks on bridges, but not analysing the bridge as a whole or from a wide perspective. The "Image Analysis and Data Fusion Technical Committee (IADF TC) of the IEEE Geoscience and Remote Sensing Society (GRSS)" created a centralized platform where researchers can find and explore datasets collected using remote sensing imagery for various applications, such as agriculture, disaster monitoring and climate change analysis (IADF TC & GRSS IEEE, 2025), and the three datasets at the top (DOTA, Bridge Dataset, AID) are taken from this platform, but they don't have damage information. This is therefore useful for an analysis of transport infrastructures too, as these are open source labelled aerial dataset (satellite view).

As these datasets have been analysed, we need to look at what technologies have been used to collect these data, understanding what is the most used one and which one is more restricted.

### 3.4 Data collection technologies used for infrastructure damage detection

The data collection technologies are presented in Table 7, where we can see that for roads, most of the datasets have been collected using normal smartphones camera, which means collecting data about roads is generally easier compared to collecting data about bridges and other transport infrastructures, and that is because any person could use their devices with camera to capture the status of the roads. In fact, the RDD dataset as we saw in Table 4, it increased from 26,336 images in the 2020 version, to 47,420 in the 2022 version, which also included more countries.

*Table 7. Data collection technologies used for damage datasets*

| Author/Year | Smartphones | Mobile mapping system | High-res cameras | Optical Device | Camera | Google Street view API |
|---|---|---|---|---|---|---|
| **Roads** | | | | | | |
| (Arya et al., 2024) | x | | x | | | x |
| (Arya, Maeda, Ghosh, Toshniwal, & Sekimoto, 2021) | x | | | | | |
| (Majidifard et al., 2020) | | | | | | x |
| (Mei & Gül, 2020) | | | | | x | |
| (F. Yang et al., 2020) | x | | | | | |
| (Stricker et al., 2019) | | x | | | | |
| (Dorafshan et al., 2018) | | | | | x | |
| (Ouma & Hahn, 2017) | x | | | | | |
| (Shi et al., 2016) | x | | | | | |
| (L. Zhang et al., 2016) | x | | | | | |
| (Oliveira & Correia, 2014) | | | | x | | |
| **Buildings** | | | | | | |
| | Smartphones | | | Satellite | | |
| (Y. Yang et al., 2024) | | | | x | | |
| (C. Wang et al., 2024) | | | | x | | |
| (V.V. et al., 2024) | | | | x | | |
| (C. Liu et al., 2022) | x | | | | | |

| | Vibration sensors | Camera | Satellite |
|---|---|---|---|
| (Gupta et al., 2019) | | | x |
| (J. Z. Xu et al., 2019) | | | x |

**Bridges**

| | Vibration sensors | Camera | Satellite |
|---|---|---|---|
| (IADF TC & GRSS IEEE, 2025) DOTA | | | x |
| (IADF TC & GRSS IEEE, 2025) Bridge Dataset | | | x |
| (IADF TC & GRSS IEEE, 2025) AID | | | x |
| (Flotzinger et al., 2023) | | x | |
| (Santaniello & Russo, 2023) | x | | |
| (H. Xu et al., 2019) | | x | |
| (Mundt et al., 2019) | | x | |

**Natural Disasters**

| | Social media | News portals | Google API | Satellite |
|---|---|---|---|---|
| (H. Chen et al., 2025) | | | | x |
| (Weber et al., 2023) | x | | x | |
| (Niloy et al., 2021) | x | x | x | |
| (Giannakeris et al., 2018) | x | | | |
| (Mouzannar et al., 2018) | x | | | |

We can observe from Table 7 that for buildings and bridges there aren't many methods for data collection, as satellites are usually the easier way to get imagery data for these infrastructures. Therefore, for buildings the data collection primarily relies on aerial images and satellite imagery, where this last one is noted for the high efficiency of capturing building damage, especially where access is restricted like in warzones. As for bridges, data collection is also limited, where technologies used are images or vibration sensor, which suggests the reliance on more specialized equipment to capture structural data and suggests the critical importance of bridge structural health as it is a more fragile infrastructure compared to roads. Here there is also data collected from satellite, but it hasn't been used for damage detection yet. In the case of Natural Disasters, data collection in this context includes the use of social media (crowdsourcing), news portals and google API. These sources are particularly useful for rapid data gathering, where for a specific study case, the data collector will most likely not be near the disaster, compared to people posting on social media and news journalists. More recently, a new dataset "BRIGHT" has been created collecting many recent natural disasters event, collected using satellite technology (H. Chen et al., 2025).

In summary, while buildings and bridges require more sophisticated equipment for data collection, for roads damages even smartphones are enough to gather data, and in the case of natural disasters, unconventional sources like social media and crowdsourcing plays an important role.

### 3.5 Types of bridge damages

In the analysis of bridge damages, we identified the common terminology used in recent studies. The primary damage types found, which are detailed in the table below, include deflection, deformation, and displacement.

*Table 8. types of bridge damage and location*

| Author/Year | Damage type | Bridge studied | Country |
| --- | --- | --- | --- |
| (Nettis et al., 2023) | Structural deformation | Albiano Magra, Fossano | Italy |
| (Yunmei et al., 2023) | Deflection | Custom | / |
| (Markogiannaki et al., 2022) | Displacement, deformation | Polyfytos | Greece |
| (Schlögl et al., 2021) | Deformation | Seitenhafenbrücke | Austria |
| (Tian et al., 2021) | Deflection | Southside of Jingtai Bridge | China |
| (Y. Wu et al., 2021) | Deflection | Custom | / |
| (Jung et al., 2019) | Deflection | Kimdaejung and Muyoung bridges | S. Korea |
| (W. Zhang et al., 2017) | Deflection | Custom | / |
| (Pan et al., 2016) | Deflection | Shuohuang railroad | China |
| (Sousa et al., 2013) | Deflection | Sorraia Bridge, Leziria Bridge | Portugal |

From Table 8, deflection is the dominant damage type, appearing in many entries of the table. This indicates that bending under load is a critical concern in bridge engineering, possibly due to heavy traffic, aging infrastructure or inadequate design. These bridges' locations indicate that damage types are not limited to specific areas and to specific bridge function, such as railroad or highway. The prevalence of deflection suggests that AI models trained on deflection-specific datasets can be effective for bridge monitoring. This can be enhanced also with InSAR, MTInSAR or D-TomoSAR, which will be mentioned in the next chapters, where they can measure minute displacements. In summary, from this table we can understand what the most common damage type is related to bridges and, more recently we see also damages labelled as displacement and deformation.

### 3.6 Methods used to detect bridge damages

Regarding the methodologies used for detecting bridge damage, we present these in the below Table 9, along with the scope and key finding from each entry. There are some different methods: satellite-based methods (MTInSAR, InSAR, D-TomoSAR), image-based methods (Digital Image Correlation), and sensor-based methods (Laser, inclinometer, vibrations responses, etc). These are also better summarized in the below Table 10.

*Table 9. Methodologies used to detect bridge damages*

| Author/Year | Method | Scope | Findings |
|---|---|---|---|
| (Corbally & Malekjafarian, 2024) | CNN framework that uses a self-calibrating Vehicle-Bridge Interaction (VBI) model to generate its own labeled training data | Classifying damage type, location, and severity | Accurately identifies the presence, type, and severity of seized bearings. Overestimates the severity of cracking and is less accurate at locating cracks at low damage levels |
| (Sarwar & Cantero, 2024) | Probabilistic Temporal Autoencoder (PTAE) using CNN and LSTM layers to analyze train-induced vibrations, paired with an EWMA control chart for damage assessment | Detecting stiffness reduction on a numerical bridge model and validating on the real KW51 railway bridge using only train-induced responses | Effectively detects subtle, progressive damage by automatically extracting features from multi-sensor data. Successfully detected structural changes on the KW51 bridge |
| (M. Huang et al., 2024) | CNN with ASAPSO hyperparameter optimization and data augmentation for scarce data scenario | Damage quantification on a lab scale continuous beam and real-world steel truss bridge | High accuracy and robust against noise, maintaining low error at a low Signal-to-Noise Ratio (SNR) |
| (Bayane et al., 2024) | Unsupervised anomaly detection algorithms | Real-time detection of an abrupt brittle cracking event | Isolation Forest and OCSVM were most robust for prompt, real-time detection of the crack. |
| (Nettis et al., 2023) | MTInSAR: Multi-Temporal satellite-based differential interferometry | Monitoring of structural deformations in bridge portfolios | Bridge with ongoing deformations have been identified and prioritized for inspection |
| (Yunmei et al., 2023) | Multi-point Chain Laser Reference | Real-time dynamic deflection detection | Measuring accuracy can reach 1 mm, and the dynamic response is good |
| (Hajializadeh, 2023) | GoogLeNet CNN using spectrograms of train-borne acceleration | Railway bridge damage detection and classification | Accurately detected and classified simulated damage with 100% accuracy using measurements from moving train. |
| Markogiannaki et al., 2022 (Markogiannaki et al., 2022) | D-TomoSAR with engineering data and forensics | Monitoring of landmark bridge | Different measurements have been taken, such as displacement products deformation trends. |
| (W. Liu et al., 2021) | Using two temporal SAR images and verifying using satellite optical image. | Damage Assessment of Bridge after flood | Four washed-away bridges were identified successfully. Three were missed due to location in radar shadow. |
| (Schlögl et al., 2021) | Time-series analysis (Persistent Scatter Interferometry) | Analysis of bridge deformation using SAR | Promising results when post-processing is correctly applied, extraction of horizontal & vertical deformations, results aggregated. Further research is needed to test transferability to other infrastructures. |
| (Tian et al., 2021) | Off axis Digital Image Correlation | Deflection measurement with Digital Image Correlation | The full-field image displacement maps can be efficiently and accurately calculated |
| (Y. Wu et al., 2021) | Secant inclination | New measurement method based on inclination of two points | Error of proposed methos is less than 1% |
| (Jung et al., 2019) | InSAR with Sentinel-1 SAR and COSMO-SkyMed data | Long-term deflection and thermal dilation of bridges | Downward movements at mid-spans, implying need for periodic monitoring |
| (W. Zhang et al., 2017) | Finite-element model with partial least-square regression | Bridge deflection estimation | The method is accurate with deflection estimation, also provides rough damage localization |
| (Pan et al., 2016) | Off-axis digital image correlation | Real-Time measurement of vertical deflection | Advanced video deflectometer is developed and can be used for field measurement of bridge deflection |
| (Sousa et al., 2013) | Strain and rotation measuements, inclinometer | Analysis of bridge deflection | On bridges, using 6$^{th}$ deg. Polynomial function, can predict vertical displacement |

Drive-by monitoring is advanced by using CNNs trained with data from self-calibrating numerical models to classify damage type and location (Corbally & Malekjafarian, 2024). For direct monitoring, approaches include using probabilistic autoencoders on train-induced vibrations to detect progressive damage (Sarwar & Cantero, 2024) or unsupervised algorithms on live sensor data to find abrupt real-world cracks (Bayane et al., 2024). To address data scarcity, another method uses data augmentation with an adaptively optimized CNN, proving effective with few-shot training samples (M. Huang et al., 2024).

MTInSAR leverages multi-temporal satellite data to detect changes over time, and similarly InSAR is applied for long term deflection and thermal dilation analysis, focusing on continuous monitoring. D-TomoSAR is the Differential Tomographic Synthetic Aperture Radar, and it's used to monitor ground deformation by analysing the differences in radar images taken at different times (M. Liu et al., 2018). A study has used two temporal SAR images to assess bridge damage due to a flood and verified the result using satellite optical imagery (W. Liu et al., 2021).

In the case of image-based methods, Digital Image Correlation and Off-axis DIC have been utilized (Tian et al., 2021) (Pan et al., 2016). This is used for deflection measurement by analyzing image displacement maps. The Off-axis DIC uses a video deflectometer to measure this.

As for sensor-based methods, an inclinometer has been used to analyze deflection using polynomial functions to predict vertical displacement (W. Zhang et al., 2017). Also, secant-inclination is also used, which measures inclination between two points to estimate deflection, achieving an error of less than 1%.

Satellite-based methods like InSAR and D-TomoSAR are valuable for inaccessible or large-scale infrastructures, which aligns with remote sensor for challenging environments such as warzones, whereas image-based offer also high-precision for specific damage types such as deflection. The data generated from these satellite-based methods can be further analysed using AI model to classify and quantify damage, which is mentioned in the next tables. The main difference between satellite-based and image-based is the time of monitoring, since methods like Multi-chain laser reference and DIC can get immediate response to structural issues, which makes them near real-time, whereas for satellite-based, some processing steps are required to be able to analyze and visualize the results, making them far from real-time, therefore more for long-term monitoring.

*Table 10. technologies used for detection of bridge damage*

| Author/Year | Type of data used | | | |
|---|---|---|---|---|
| | **Media** | **Sensor** | **SAR** | **Laser** |
| (Corbally & Malekjafarian, 2024) | | Accelerometer | | |
| (Sarwar & Cantero, 2024) | | SHM system | | |
| (M. Huang et al., 2024) | | Accelerometer | | |
| (Bayane et al., 2024) | | Accelerometer | | |
| (Nettis et al., 2023) | | | MTInSAR | |
| (Yunmei et al., 2023) | | | | Chain Laser beam |
| (Markogiannaki et al., 2022) | | | D-TomoSAR | |
| (W. Liu et al., 2021) | | | SAR | |
| (Schlögl et al., 2021) | | | SAR | |
| (Tian et al., 2021) | Video deflectometer side of bridge | | | Rangefinder |
| (Jung et al., 2019) | | | InSAR | |
| (W. Zhang et al., 2017) | | Inclinometer | | |
| (Pan et al., 2016) | Video deflectometer side of bridge | | | Rangefinder |
| (Sousa et al., 2013) | | Inclinometer | | |

## 3.7 Applications of satellite data methods and Synthetic Aperture Radar (SAR)

Satellite imagery and Synthetic Aperture Radar (SAR) have been analysed and seen in the previous tables, however this Table 11 summarises some applications of SAR and the integration with AI where possible. The table is divided into three sections, including General SAR applications, SAR with Coherence and long-term monitoring.

*Table 11. applications of satellite SAR methods and uses of AI models*

| Author/Year | Application | AI | Satellite |
|---|---|---|---|
| (C. Wang et al., 2024) | Building damage assessment | Bayesian Networks | Maxar Sentinel-1 |
| (Markogiannaki et al., 2022) | Monitoring of a landmark bridge | No | Sentinel-1A/B |
| (X. Huang et al., 2022) | Marine oil spill detection | Faster R-CNN | Sentinel-1 Radarsat-2 |
| (Heiselberg, 2020) | Ship-Iceberg classification (multispectral images) | SVM & CNN | Sentinel-1 Sentinel-2 |
| (R. Wu et al., 2020) | Mapping glacial lakes (with optical satellite) | CNN | Landsat 8 (opt) Sentinel-1A |
| (Nemni et al., 2020) | Rapid flood segmentation | FCNN | Sentinel-1 |
| (Winsvold et al., 2018) | Regional glacier mapping | No | Sentinel-1A Radarsat-2 |
| (Henry et al., 2018) | Road segmentation in satellite images | FCNN | TerraSAR-X |
| (Rahman & Thakur, 2018) | Detection, mapping and analysis of flood propagation with GIS | No | Radarsat |
| (Markert et al., 2018) | Surface water mapping (with optical satellite) | No | Sentinel-1 Landsat (opt) |
| (Chang et al., 2017) | Nationwide Railway monitoring | No | Radarsat-2 |
| | **With Coherence product** | | |
| (Kopiika et al., 2025) | Rapid post-disaster infrastructure damage characterization enabled by remote sensing and deep learning technologies | SAM (Segment Anything Model) | Maxar Sentinel-1 |
| (Y. Yang et al., 2024) | Building damage assessment | RNN | Sentinel-1 |
| (Lopez-Sanchez et al., 2023) | Multi-Annual Evaluation of Time Series of Sentinel-1 Inter. Coherence as a tool for Crop Monitoring | No | Sentinel-1 |
| (ElGharbawi & Zarzoura, 2021) | Damage detection using SAR coherence statistical analysis, application to Beirut, Lebanon | No | Sentinel-1 |
| (Sun et al., 2020) | Deep Learning Framework for SAR Interferometric Phase Restoration and Coherence Estimation | CNN | TerraSAR-X |
| (Sharma et al., 2017) | Earthquake Damage Visualization for Rapid Detection of Earthquake-Induced damage | No | JAXA ALOS-2 |
| (Yun et al., 2015) | Rapid Damage Mapping for 2015 Gorkha Earthquake | No | COSMO-SkyMed, JAXA ALOS-2 |
| (Bouaraba et al., 2012) | Detection of surface changes using Coherence Change Detection | No | COSMO-SkyMed |
| (Preiss et al., 2006) | Detection of scene changes with Change in Coherence | No | DSTO Ingara X-Band SAR |
| **Long Term Monitoring** | | | |
| (Tonelli et al., 2023) | Interpretation of Bridge Health Monitoring Data from Satellite InSAR | No | COSMO-SkyMed Sentinel-1 |
| (Nettis et al., 2023) | Multi-Temporal satellite-based differential interferometry for monitoring structural deformations of bridge portfolios | No | COSMO-SkyMed |
| (Jung et al., 2019) | Long-Term Deflection Monitoring for Bridges Using X and C-Band Time-Series SAR Interferometry | No | COSMO-SkyMed |

In the first section it's presented how SAR is useful when it comes to detecting marine oil spills, ship-iceberg detection, glacial lake mappings, road segmentation and water/flood mapping. Here the satellites that have been used include two missions from ESA (European Space Agency), Sentinel-1 and Sentinel-2, TerraSAR-X, Landsat and Radarsat. Some of these cases have utilised AI models, such as Faster R-CNN, Support Vector Machine (SVM) and Convolutional Networks for automated detection and classification (X. Huang et al., 2022) (Heiselberg, 2020) (R. Wu et al., 2020) (Nemni et al., 2020) (Henry et al., 2018).

In SAR interferometry, coherence indicates a measure of correlation between two SAR images at different times, where high coherence indicates better interferences and therefore more accurate phase measurements (Y. Zhang & Prinet, 2004). This is here used for rapid-post disaster infrastructure damage characterization (Kopiika et al., 2025), crop monitoring (Lopez-Sanchez et al., 2023), earthquake damage visualization (Sharma et al., 2017) (Yun et al., 2015) and scene change (ElGharbawi & Zarzoura, 2021) (Bouaraba et al., 2012) (Preiss et al., 2006). The coherence product is mainly taken from Sentinel-1 mission, but also from the German TerraSAR-X, the Japanese JAXA ALOS-2 and the Italian COSMO-SkyMed mission (see Table 12 for available satellites used for monitoring infrastructures along with more specifics). Some AI models have been used here as well, but less frequent compared to general SAR application. In this case, SAM (Segment Anything Model) and CNN are used for tasks like phase restoration and coherence estimation. Therefore, coherence product can be highly useful when comparing pre- and post-event SAR images.

Lastly, for Long-term monitoring, there are two cases of bridge health monitoring and multi-temporal monitoring of structural deformations, using mainly Sentinel-1 and COSMO-SkyMed, without any case of using AI models.

The table shows the versatility of SAR and its usage across different domains. Also, AI integration shows the potential of machine learning to automate and scale SAR data analysis. The frequent use of Sentinel-1 mission from ESA shows the accessibility of high-quality radar imagery which is crucial for researchers. We also saw how the coherence product can be invaluable for post disaster assessment in challenging environments (Kopiika et al., 2025).

The absence of AI usage for long-term monitoring suggests a gap in utilising Machine Learning for continuous infrastructure monitoring, possibly due to the fact that SAR requires long processing times and expertise.

Therefore, while SAR offers unique advantages for infrastructure monitoring, it has some challenges, as mentioned above. The complexity of SAR data that arises from the multiple dimensions, polarizations and frequency, impacts image resolution, sensitivity to surface features and penetration depth. Atmospheric conditions also further complicate it, with effects such as attenuation, ionospheric disturbances, and tropospheric distortions leading to signal loss and reducing image quality. It is also hard to interpret, due to its signal noise, speckle, distortion and scattering effects, presented in grayscale which requires advanced training (Deep Block, 2023).

*Table 12. Available satellite data for monitoring infrastructures. GSD: Ground Sample Distance*

| Author | Satellite data source | Data resolution in GSD | Features |
|---|---|---|---|
| (Gupta et al., 2019) | Maxar | 0.3m | Assessing building damages after natural disasters. MDA. |
| (Mari et al., 2018) | COSMO-SkyMed | 1m - 100m | High resolution imagery, multi-mode operation and dual polarization capability. Italian Space Agency. |
| (Motohka et al., 2017) | JAXA ALOS-2 | 1m x 3m (spotlight), 3m,6m,10m (strimap) | High resolution imagery, L-band SAR, Compact InfraRed Camera, Automatic Ship Identification System. Japanese Aerospace Exploration Agency. |
| (Chabot et al., 2014) | RADARSAT-2 | 3m – 100m | High resolution imaging, flexible polarization and left/right looking imaging capabilities. C-Band SAR. Canadian Space Agency. |
| (Roy et al., 2014) | Landsat 7/8 | 15/30m | Landsat 8 has narrower spectral bands, improved calibration and signal-to-noise characteristics, high radiometric resolution and more precise geometry compared to Landsat 7. NASA and US. |
| (Space Agency, 2012a) | Sentinel-1, ESA | 5m - 40m | Radar imagery, dual polarization, short revisit times, fast product delivery. ESA. |
| (Space Agency, 2012b) | Sentinel-2, ESA | 10m | Wide-swath, high resolution and multi spectral imager for earth surface monitoring. ESA. |
| (Werninghaus & Buckreuss, 2010) | TerraSAR-X | 1m - 40m | Radar imagery, various imaging modes, high resolution, rapid switching between modes and polarizations. German Aerospace Centre and Airbus. |

Some of the studies above have used damage quantification methods, which have been listed here in Table 13, highlighting their application in real-world scenarios for assessing infrastructure damage, especially in the context of natural disasters. We can see there are 3 methods used for roads infrastructures, such as PASER, PCI and SDI which are standardized visual survey methods that are crucial for systematic infrastructure maintenance planning. Methods like Hazus and UNOST are relevant for post disaster assessment.

*Table 13. Damage quantification methodologies*

| Author/Year | Damage quantification methods | Data source used | Infrastructure | Case study applications |
|---|---|---|---|---|
| (C. Wang et al., 2024) | StEER network | Visual survey, NOAA, Maxar, Copernicus | Natural disasters | Hurricane Ian |
| (Teopilus & Amrozi, 2023) | PASER | Visual survey, Bina Marga | Roads | Dandels road, Java island |
| (Teopilus & Amrozi, 2023) | PCI | Visual survey, Bina Marga | Roads | Dandels road, Java island |
| (Teopilus & Amrozi, 2023) | SDI | Visual survey, Bina Marga | Roads | Dandels road, Java island |
| (Gupta et al., 2019) | Hazus Fema | Maxar | Natural disasters | xView2 competition |
| (J. Z. XU ET AL., 2019) | UNOSAT | UNITAR | Buildings | Indonesia 2018, Mexico City 2017, Haiti 2010 |

## 3.8 Natural hazards in studies using AI models

Having detailed the specific damages, AI models, and data technologies, the focus now shifts to the broader context of the causal events. The type of natural hazards, such as flood, earthquake, or wildfire, directly influences the nature and scale of damage to transport infrastructure. This link is critical, as the hazard determines the most suitable remote sensing data and consequently, the design and application of AI models for assessment. To understand the current state of research from this perspective, the following sections analyse the specific natural hazards that have been the focus of using AI models.

The following table summarizes a selection of studies that identify the specific natural hazard stressors investigated. This illustrates the area of focus within the scientific community regarding use of AI for disaster management and risk assessment.

*Table 14. Natural hazards stressors analysed*

| Author/Year | Wild-Fire | Flood | Land disaster | Nature disaster | Earthquake | Hurricane | Volcano | Human-induced | Tsunami |
|---|---|---|---|---|---|---|---|---|---|
| (Bhardwaj et al., 2025) | | | | | | x | | | |
| (H. Chen et al., 2025) | | x | | | x | x | | x | x |
| (Braik & Koliou, 2024) | | | | | | x | | | |
| (Y. Yang et al., 2024) | | | | | x | | | | |
| (C. Wang et al., 2024) | | | | | | x | | | |
| Weber et al., 2022 (Weber et al., 2022) | x | x | x | | x | x | x | x | |
| (Niloy et al., 2021) | x | x | x | | | | | x | x |
| Arif et al., 2020 (Arif et al., 2020) | x | x | | x | | | | x | |
| (Gupta et al., 2019) | x | x | | | x | x | x | | x |
| (Rizk et al., 2019) | | | | x | | | | | |
| (Barz et al., 2018) | | x | | | | | | | |
| Giannakeris et al., 2018 (Giannakeris et al., 2018) | x | x | | | | | | | |
| (Mouzannar et al., 2018) | x | x | | x | | | | x | |

| | | | | |
|---|---|---|---|---|
| Muhammad et al., 2017 (Muhammad et al., 2017) | x | | | |
| (Alam et al., 2017) | | x | x | x |

The hazards listed in Table 14 directly threaten the integrity of transport infrastructures, which are vital for economic and social connectivity. The table suggests that most of these natural hazards are floods and hurricanes, which can cause direct damage to roads and bridges, and as well as indirect impacts through cascading events like traffic disruptions. As presented in Table 13 a range of methods have been employed for damage quantification. In the context of natural disasters, methodologies such as StEER network (C. Wang et al., 2024), HAZUS (developed by FEMA) and UNOSAT are frequently adopted. Specifically, HAZUS-FEMA has been applied for multi-hazard damage classification, encompassing events such as floods, hurricanes, and earthquakes (Gupta et al., 2019). These tools leverage geospatial data and standardized assessment protocols to support large-scale disaster impact evaluation.

The table also includes entries for human-induced hazards, which connects to the discussion of using remote sensing in challenging situations such as warzones. Technologies such as SAR can operate in these areas and are valuable for assessing infrastructure damage in these contexts, as in (Kopiika et al., 2025). The constraint here is the suitability of methodologies and data in these contexts, i.e. the suitability of AI model for SAR based monitoring is limited by the slow acquisition and processing of SAR data, and therefore in the case of rapid-onset hazards like the ones above, timely damage assessment is critical. In Table 15 and Table 16 the datasets and AI models used are presented.

A further constraint is the resolution of the satellite technologies, where open-source satellite missions have worse resolution compared to commercial satellites, such as MAXAR, which can achieve a resolution of 0.3m GSD (Ground Sample Distance), as presented in Table 12. This has an impact on the accessibility of resources, especially in the field of research where these are limited.

The suitability of the technologies shown so far is constrained by practical challenged mentioned before, like data unreadiness, need for faster processing, and better access to high-quality resources. Addressing these gaps through extended datasets and innovative processing techniques will be crucial for advancing infrastructure resilience, in line with supporting sustainable and climate-aware transport networks.

*Table 15. Available natural disaster datasets*

| Author/Year | Dataset Name | Classes | Size | Geo area |
|---|---|---|---|---|
| (H. Chen et al., 2025) | BRIGHT | Tsunami, Hurricane, Flood, Earthquake, Human-induced | 4,246 | Lebanon, Equatorial Guinea, Congo, Haiti, Spain, USA, Ukraine, Turkey, Myanmar, Morocco, Libya, Mexico, Japan |
| (C. Wang et al., 2024) | StEER Hurrican IAN | No damage, Minor, Moderate, Severe, Destroyed | 2,472 | USA (Florida) |
| (Weber et al., 2023) | Incidents1M | 43 | 977,088 | Worldwide |
| (Niloy et al., 2021) | Disaster-Dataset | Fire, Water, Infrastructure, human damage, land disaster, non-damage. | 13,720 | India, Japan, Australia, California, Brazil |
| (Arif et al., 2020) | SAD | Fire, Flood, Infrastructure, Nature, Human damage, non-damage. | 493 | South Asia |
| (Barz et al., 2018) | EU-Flood | Flooding, Inundation depth, water pollution. | 3,435 | Europe |
| (Rizk et al., 2019) | Home-grown + Sun dataset | Infrastructure, Natural disaster. | 2,344 | Nepal, Chile, Japan, Kenya |
| (Giannakeris et al., 2018) | 3F-emergency dataset | Fire, Flood. | 12,000 | N/a |
| (Mouzannar et al., 2018) | UCI | Fire, Flood, Infrastructure, Nature, Human damage, non-damage. | 5,880 | Worldwide |
| (Muhammad et al., 2017) | (Chino et al., 2015) (Foggia et al., 2015) (Verstockt et al., 2013) (Ko et al., 2011) | Fire, non-damage. | 68,457 | N/a |
| (Alam et al., 2017) | Image4act | Earthquakes, Typhoon, Hurricane. | 34,562 | Nepal, Ecuador, Philippines, Haiti |

Table 15 presents datasets regarding natural hazards that included use of AI model for detection, and it varies significantly in scale and scope, reflecting the diversity of natural hazards impacting infrastructures. The most recent one, named "BRIGHT", has a collection of labelled data for damaged buildings from recent natural disasters (H. Chen et al., 2025). Incidents1M stands out as the largest one, however it included many classes not related to natural disasters such as "bus accident" "motorcycle accident" and other similar accidents, but apart from this it contains a large number of natural disaster classes, such as "dust devil", "heavy rainfall", "storm surge" and so on (Weber et al., 2023). The scale of this datasets makes it ideal for AI training. In contrast, other datasets such as SAD (Arif et al., 2020) and Home-grown + Sun dataset (Rizk et al., 2019) are more regionally focused, limiting applicability. These datasets complement the remote sensing technologies, such as SAR and Sentinel-1 previously discussed, i.e. the EU-flood dataset (Barz et al., 2018) aligns with the possibility of using SAR for flood detection (segmentation).

*Table 16. Previously used AI models for damage detection after natural disaster*

| Author/Year | Dataset | Satellite | Model |
| --- | --- | --- | --- |
| (Braik & Koliou, 2024) | xBD | Yes | CNN |
| (C. Wang et al., 2024) | STEER | Yes | Bayesian Networks |
| (Kaur et al., 2023) | xBD, Ida-BD, LEVIR-CD | Yes | Transformer |
| (Weber et al., 2023) | Incidents1M | No | ResNet50 |
| (C. Wu et al., 2021) | xBD + Maxar | Yes | Attention U-Net |
| (Gupta & Shah, 2020) | xBD | Yes | RescueNet |
| (Arif et al., 2020) | SAD | No | VGG16 |
| (Weber & Kané, 2020) | xBD | Yes | Mask R-CNN |
| (Bai et al., 2020) | xBD | Yes | PPM-SSNet |
| (Potnis et al., 2019) | WorldView-2 | Yes | ERFNet |
| (Mouzannar et al., 2018) | Home-grown | No | DFMC with SVM |
| (Alam et al., 2017) | Image4act | No | VGG16 |

After discussing the datasets, we analyse also the AI model that have been used to achieve the scope, and in Table 16 these are displayed, with information about whether satellite technology have been used and the specific AI model. These range from traditional deep learning architectures such as ResNEt50, VGG16, to more specialized for specific scenarios, like RescueNet and Attention U-Net. Notice the frequent use of xBD dataset (Gupta et al., 2019), which underscore its importance in building damage assessment, due to its extensive satellite imagery (22,068) and standardized damage classification using Hazus FEMA.

The choice of models reflects their suitability for specific tasks. For example, Attention U-Net (C. Wu et al., 2021) and Mask R-CNN (Weber & Kané, 2020), were used with xBD for segmentation tasks, identifying damaged areas in satellite imagery. On the other hand, VGG16 (Arif et al., 2020) and ResNet50 (Weber et al., 2023) are more general purpose as they haven't used satellite data and focused on simpler classification task rather than fine-grained damage mapping.

The reliance on satellite data in AI applications highlights the practical challenged discussed earlier, such as slow processing of SAR data as mentioned in section 3.6. While optical satellite imagery offers high resolution (see Table 12) it is weather dependent, limiting the effectiveness during events such as hurricanes and floods. Therefore, SAR overcomes this issue but requires post-processing, which delays the damage assessment for these natural hazards.

## 4. Conclusion

Transport infrastructures are essential to the vitality of modern economies and societies, yet they are still vulnerable to impacts of climate change and natural disasters. Therefore, the demand for rapid damage assessment and monitoring systems is more urgent than ever. In this paper, we examined the transformative role of emerging digital technologies, focusing on AI models and on remote sensing (satellite technology) in strengthening the resilience of transport infrastructures, such as roads and bridges, and with focus also on buildings. The potential of these technologies, although remains constrained by practical and data-related challenges.

In this review, we highlighted the AI models and datasets used for different infrastructures. A key finding is the significant disparity in research focus: while data and models for road damage detection are abundant, reflecting the ease of data acquisition, there is a distinct scarcity of studies integrating AI with SAR data for comprehensive bridge damage assessment. Although models such as ResNet50, Attention U-Net and Mask R-CNN show promise, there is still lack of comparative studies especially for satellite imagery-based approaches, and therefore their effectiveness across varied contexts is not yet fully understood.

SAR technology with its capabilities and variants (i.e. MTInSAR and D-TomoSAR), it excels in monitoring structural deformation with high precision, however it is still limited by complex data structures, atmospheric distortions, interpretive challenges and big computational demand. Some initiatives, such as AI4SAR (ICEYE OY (FI), n.d.), are designing solutions by leveraging AI to streamline SAR data processing, hoping to offer more accessible and efficient monitoring.

Some key direction to advance the field includes:

- Comparative research of newer AI models to determine the most effective solutions for different infrastructure types and hazards, with emphasis on remote technologies such as satellite imagery
- Expand datasets to include underrepresented classes (hazards, infrastructure categories)
- Multi-sensor integration that merges SAR, optical-imagery and ground-based sensors for a complete assessment of infrastructure health
- Use of AI to optimize SAR data analysis and reducing computational barriers, moving towards near-real-time monitoring
- Explore AI-driven approaches for continuous infrastructure monitoring, especially for critical ones such as bridges.

In conclusion, this review covered the latest technologies, including latest AI models and datasets used for damage assessment for various transport infrastructures. Furthermore, we analysed the use of remote technologies, such as satellites, for data acquisition. However, these technologies are constrained by some limitations, as we saw above. As noted in a comprehensive review on data readiness for AI, poor quality data can compromise AI model accuracy, a challenge relevant to the complicated and unstructured nature of SAR data for example and mentioned that while metrics for assessing data readiness for AI are advancing, standardized approaches remain underdeveloped (Hiniduma et al., 2025). Initiative like AI4SAR demonstrate progress in leveraging SAR effectively, yet future research must prioritize not only technological advancement, but also robust and standardized metrics for evaluating data readiness specific to transport infrastructures. This will ensure AI driven solutions deliver efficient, reliable and sustainable outcomes.

**Acknowledgments**

This research received funding by the UK Research and Innovation (UKRI) under the UK government's Horizon Europe funding guarantee [grant agreement No: EP/Y003586/1, EP/X037665/1]. This is the funding guarantee for the European Union HORIZON-MSCA-2021-SE-01 [grant agreement No: 101086413] ReCharged - Climate-aware Resilience for Sustainable Critical and interdependent Infrastructure Systems enhanced by emerging Digital Technologies.

**References**

*A pro-innovation approach to AI regulation*. (2023). 91. https://www.gov.uk/government/publications/ai-regulation-a-pro-innovation-approach

Abduljabbar, R., Dia, H., Liyanage, S., & Bagloee, S. A. (2019). Applications of Artificial Intelligence in Transport: An Overview. *Sustainability 2019, Vol. 11, Page 189*, *11*(1), 189. https://doi.org/10.3390/SU11010189

Abedi, M., Shayanfar, J., & Al-Jabri, K. (2023). Infrastructure damage assessment via machine learning approaches: a systematic review. *Asian Journal of Civil Engineering*, *24*(8), 3823–3852. https://doi.org/10.1007/S42107-023-00748-5/TABLES/7

Abubakr, M., Rady, M., Badran, K., & Mahfouz, S. Y. (2024). Application of deep learning in damage classification of reinforced concrete bridges. *Ain Shams Engineering Journal*, *15*(1), 102297. https://doi.org/10.1016/J.ASEJ.2023.102297

Agbaje, T. H., Abomaye-Nimenibo, N., Ezeh, C. J., Bello, A., & Olorunnishola, A. (2024). Building Damage Assessment in Aftermath of Disaster Events by Leveraging Geoai (Geospatial Artificial Intelligence): Review. *Https://Wjarr.Co.in/Sites/Default/Files/WJARR-2024-2000.Pdf*, *23*(1), 667–687. https://doi.org/10.30574/WJARR.2024.23.1.2000

Alam, F., Imran, M., & Ofli, F. (2017). Image4Act: Online social media image processing for disaster response. *Proceedings of the 2017 IEEE/ACM International Conference on Advances in Social Networks Analysis and Mining, ASONAM 2017*, 601–604. https://doi.org/10.1145/3110025.3110164

Angulo, A., Vega-Fernández, J. A., Aguilar-Lobo, L. M., Natraj, S., & Ochoa-Ruiz, G. (2019). Road Damage Detection Acquisition System Based on Deep Neural Networks for Physical Asset Management. *Lecture Notes in Computer Science (Including Subseries Lecture Notes in Artificial Intelligence and Lecture Notes in Bioinformatics)*, *11835 LNAI*, 3–14. https://doi.org/10.1007/978-3-030-33749-0_1/TABLES/5

Argyroudis, S. A., Mitoulis, S. A., Chatzi, E., Baker, J. W., Brilakis, I., Gkoumas, K., Vousdoukas, M., Hynes, W., Carluccio, S., Keou, O., Frangopol, D. M., & Linkov, I. (2022). Digital technologies can enhance climate resilience of critical infrastructure. *Climate Risk Management*, *35*, 100387. https://doi.org/10.1016/J.CRM.2021.100387

Arif, Omar, A., Ashraf, S., Rahman, A. K. M. M., Amin, M. A., & Ali, A. A. (2020). A comparative study on disaster detection from social media images using deep learning. *Advances in Intelligent Systems and Computing*, *1112*, 485–499. https://doi.org/10.1007/978-981-15-2188-1_38/COVER

*Artificial Intelligence Act: MEPs adopt landmark law | News | European Parliament*. (2024, March 13). https://www.europarl.europa.eu/news/en/press-room/20240308IPR19015/artificial-intelligence-act-meps-adopt-landmark-law

Arya, D., Maeda, H., Ghosh, S. K., Toshniwal, D., Mraz, A., Kashiyama, T., & Sekimoto, Y. (2021). Deep learning-based road damage detection and classification for multiple countries. *Automation in Construction*, *132*, 103935. https://doi.org/10.1016/J.AUTCON.2021.103935

Arya, D., Maeda, H., Ghosh, S. K., Toshniwal, D., Omata, H., Kashiyama, T., & Sekimoto, Y. (2022). Crowdsensing-based Road Damage Detection Challenge (CRDDC'2022).

*Proceedings - 2022 IEEE International Conference on Big Data, Big Data 2022*, 6378–6386. https://doi.org/10.1109/BIGDATA55660.2022.10021040

Arya, D., Maeda, H., Ghosh, S. K., Toshniwal, D., & Sekimoto, Y. (2021). RDD2020: An annotated image dataset for automatic road damage detection using deep learning. *Data in Brief*, *36*, 107133. https://doi.org/https://doi.org/10.1016/j.dib.2021.107133

Arya, D., Maeda, H., Sekimoto, Y., Omata, H., Ghosh, S. K., Toshniwal, D., Sharma, M., Pham, V. V., Zhong, J., Al-Hammadi, M., Shami, M. B., Nguyen, D., Cheng, H., Zhang, J., Klein-Paste, A., Mork, H., Lindseth, F., Seto, T., Mraz, A., & Kashiyama, T. (2024). RDD2022: A multi-national image dataset for automatic road damage detection. *Geoscience Data Journal*, *11*, 846–862. https://doi.org/10.1002/GDJ3.260

Bai, Y., Hu, J., Su, J., Liu, X., Liu, H., He, X., Meng, S., Mas, E., & Koshimura, S. (2020). Pyramid Pooling Module-Based Semi-Siamese Network: A Benchmark Model for Assessing Building Damage from xBD Satellite Imagery Datasets. *Remote Sensing 2020, Vol. 12, Page 4055*, *12*(24), 4055. https://doi.org/10.3390/RS12244055

Barz, B., Schröter, K., Münch, M., Yang, B., Unger, A., Dransch, D., Denzler, J., & E -p R I N T, P. R. (2018). Enhancing Flood Impact Analysis using Interactive Retrieval of Social Media Images. *Archives of Data Science, Series A (Online First)*, *5*. https://doi.org/10.5445/KSP/1000087327/06

Bayane, I., Leander, J., & Karoumi, R. (2024). An unsupervised machine learning approach for real-time damage detection in bridges. *Engineering Structures*, *308*, 117971. https://doi.org/10.1016/J.ENGSTRUCT.2024.117971

Bhardwaj, D., Nagabhooshanam, N., Singh, A., Selvalakshmi, B., Angadi, S., Shargunam, S., Guha, T., Singh, G., & Rajaram, A. (2025). Enhanced satellite imagery analysis for post-disaster building damage assessment using integrated ResNet-U-Net model. *Multimedia Tools and Applications*, *84*(5), 2689–2714. https://doi.org/10.1007/S11042-024-20300-0/FIGURES/9

Bouaraba, A., Younsi, A., Belhadj-Aissa, A., Acheroy, M., Milisavljevic, N., & Closson, D. (2012). Robust techniques for coherent change detection using COSMO-SkyMed SAR images. *Progress In Electromagnetics Research M*, *22*, 219–232. https://doi.org/10.2528/PIERM11110707

Braik, A. M., & Koliou, M. (2024). Automated building damage assessment and large-scale mapping by integrating satellite imagery, GIS, and deep learning. *Computer-Aided Civil and Infrastructure Engineering*, *39*(15), 2389–2404. https://doi.org/10.1111/MICE.13197

Chabot, M., Decoust, C., Ledantec, P., Williams, D., Hillman, A., Rolland, P., & Periard, R. (2014). RADARSAT-2 system operations and performance. *International Geoscience and Remote Sensing Symposium (IGARSS)*, 994–997. https://doi.org/10.1109/IGARSS.2014.6946594

Chang, L., Dollevoet, R. P. B. J., & Hanssen, R. F. (2017). Nationwide Railway Monitoring Using Satellite SAR Interferometry. *IEEE Journal of Selected Topics in Applied Earth Observations and Remote Sensing*, *10*(2), 596–604. https://doi.org/10.1109/JSTARS.2016.2584783

Chen, H., Song, J., Dietrich, O., Broni-Bediako, C., Xuan, W., Wang, J., Shao, X., Wei, Y., Xia, J., Lan, C., Schindler, K., & Yokoya, N. (2025). BRIGHT: A globally distributed multimodal building damage assessment dataset with very-high-resolution for all-weather disaster response. *Earth System Science Data*. https://doi.org/10.5194/ESSD-2025-269

Chen, J., Yu, X., Li, Q., Wang, W., & He, B.-G. (2024). LAG-YOLO: Efficient road damage detector via lightweight attention ghost module. *Journal of Intelligent Construction*, *2*(1), 9180032. https://doi.org/10.26599/JIC.2023.9180032

Chino, D. Y. T., Avalhais, L. P. S., Rodrigues, J. F., & Traina, A. J. M. (2015). BoWFire: Detection of Fire in Still Images by Integrating Pixel Color and Texture Analysis. *2015 28th SIBGRAPI Conference on Graphics, Patterns and Images*, 95–102. https://doi.org/10.1109/SIBGRAPI.2015.19

Corbally, R., & Malekjafarian, A. (2024). A deep-learning framework for classifying the type, location, and severity of bridge damage using drive-by measurements. *Computer-Aided Civil and Infrastructure Engineering*, *39*(6), 852–871. https://doi.org/10.1111/MICE.13104;CTYPE:STRING:JOURNAL

de Abreu, V. H. S., Santos, A. S., & Monteiro, T. G. M. (2022). Climate Change Impacts on the Road Transport Infrastructure: A Systematic Review on Adaptation Measures. *Sustainability 2022, Vol. 14, Page 8864*, *14*(14), 8864. https://doi.org/10.3390/SU14148864

Deep Block. (2023, March 27). How AI can help overcome SAR imagery analysis challenges. | LinkedIn. *The Deep Dive*. https://www.linkedin.com/pulse/how-ai-can-help-overcome-sar-imagery-analysis-challenges/

Díaz-Rodríguez, N., Del Ser, J., Coeckelbergh, M., López de Prado, M., Herrera-Viedma, E., & Herrera, F. (2023). Connecting the dots in trustworthy Artificial Intelligence: From AI principles, ethics, and key requirements to responsible AI systems and regulation. *Information Fusion*, *99*, 101896. https://doi.org/10.1016/J.INFFUS.2023.101896

Dorafshan, S., Thomas, R. J., & Maguire, M. (2018). SDNET2018: An annotated image dataset for non-contact concrete crack detection using deep convolutional neural networks. *Data in Brief*, *21*, 1664–1668. https://doi.org/10.1016/J.DIB.2018.11.015

Du, Y., Pan, N., Xu, Z., Deng, F., Shen, Y., & Kang, H. (2021). Pavement distress detection and classification based on YOLO network. *International Journal of Pavement Engineering*, *22*(13), 1659–1672. https://doi.org/10.1080/10298436.2020.1714047

ElGharbawi, T., & Zarzoura, F. (2021). Damage detection using SAR coherence statistical analysis, application to Beirut, Lebanon. *ISPRS Journal of Photogrammetry and Remote Sensing*, *173*, 1–9. https://doi.org/10.1016/J.ISPRSJPRS.2021.01.001

*European Space Agency*. (2025). https://www.esa.int/

Flotzinger, J., Rösch, P. J., & Braml, T. (2023). *dacl10k: Benchmark for Semantic Bridge Damage Segmentation*. 8611–8620. https://doi.org/10.1109/wacv57701.2024.00843

Foggia, P., Saggese, A., & Vento, M. (2015). Real-Time Fire Detection for Video-Surveillance Applications Using a Combination of Experts Based on Color, Shape, and Motion. *IEEE Transactions on Circuits and Systems for Video Technology*, *25*(9), 1545–1556. https://doi.org/10.1109/TCSVT.2015.2392531

Gao, Y., Li, H., & Fu, W. (2023). Few-shot learning for image-based bridge damage detection. *Engineering Applications of Artificial Intelligence*, *126*, 107078. https://doi.org/10.1016/J.ENGAPPAI.2023.107078

Giannakeris, P., Avgerinakis, K., Karakostas, A., Vrochidis, S., & Kompatsiaris, I. (2018). People and Vehicles in Danger - A Fire and Flood Detection System in Social Media. *2018 IEEE 13th Image, Video, and Multidimensional Signal Processing Workshop (IVMSP)*, 1–5. https://doi.org/10.1109/IVMSPW.2018.8448732

Guo, G., & Zhang, Z. (2022). Road damage detection algorithm for improved YOLOv5. *Scientific Reports 2022 12:1*, *12*(1), 1–12. https://doi.org/10.1038/s41598-022-19674-8

Gupta, R., Goodman, B., Patel, N. N., Hosfelt, R., Sajeev, S., Heim, E. T., Doshi, J., Lucas, K., Choset, H., & Gaston, M. E. (2019). xBD: A Dataset for Assessing Building Damage from Satellite Imagery. *ArXiv*, *abs/1911.09296*. https://api.semanticscholar.org/CorpusID:198167037

Gupta, R., & Shah, M. (2020). RescueNet: Joint building segmentation and damage assessment from satellite imagery. *Proceedings - International Conference on Pattern Recognition*, 4405–4411. https://doi.org/10.1109/ICPR48806.2021.9412295

Hajializadeh, D. (2023). Deep learning-based indirect bridge damage identification system. *Structural Health Monitoring*, *22*(2), 897–912. https://doi.org/10.1177/14759217221087147/ASSET/2B24CECC-FA14-412A-9CC5-A3DF70AC2D2C/ASSETS/IMAGES/LARGE/10.1177_14759217221087147-FIG13.JPG

Hegde, V., Trivedi, D., Alfarrarjeh, A., Deepak, A., Ho Kim, S., & Shahabi, C. (2020). Yet Another Deep Learning Approach for Road Damage Detection using Ensemble Learning. *Proceedings - 2020 IEEE International Conference on Big Data, Big Data 2020*, 5553–5558. https://doi.org/10.1109/BigData50022.2020.9377833

Heiselberg, H. (2020). Ship-Iceberg Classification in SAR and Multispectral Satellite Images with Neural Networks. *Remote Sensing 2020, Vol. 12, Page 2353*, *12*(15), 2353. https://doi.org/10.3390/RS12152353

Hendrycks, D., Woodside, T., & Mazeika, M. (2023). *An Overview of Catastrophic AI Risks*.

Henry, C., Azimi, S. M., & Merkle, N. (2018). Road segmentation in SAR satellite images with deep fully convolutional neural networks. *IEEE Geoscience and Remote Sensing Letters*, *15*(12), 1867–1871. https://doi.org/10.1109/LGRS.2018.2864342

Hiniduma, K., Byna, S., & Bez, J. L. (2025). Data Readiness for AI: A 360-Degree Survey. *ACM Computing Surveys*, *57*(9), 1–39. https://doi.org/10.1145/3722214

Huang, M., Zhang, J., Li, J., Deng, Z., & Luo, J. (2024). Damage identification of steel bridge based on data augmentation and adaptive optimization neural network. *Structural Health Monitoring*. https://doi.org/10.1177/14759217241255042/ASSET/94750AE5-6CFD-4FDB-84C0-83A1038CA580/ASSETS/IMAGES/LARGE/10.1177_14759217241255042-FIG20.JPG

Huang, X., Zhang, B., Perrie, W., Lu, Y., & Wang, C. (2022). A novel deep learning method for marine oil spill detection from satellite synthetic aperture radar imagery. *Marine Pollution Bulletin*, *179*, 113666. https://doi.org/10.1016/J.MARPOLBUL.2022.113666

IADF TC, & GRSS IEEE. (2025). *Earth Observation*. https://eod-grss-ieee.com/dataset-search

ICEYE OY (FI). (n.d.). *Artificial Intelligence for SAR at High Resolution (AI4SAR HighRes) - eo science for society*. Eo Science for Society. Retrieved 6 May 2025, from https://eo4society.esa.int/projects/ai4sar-highres/

Janjeva, A., Gausen, A., Mercer, S., & Sippy, T. (2024). *Evaluating Malicious Generative AI Capabilities: Understanding inflection points in risk*. https://cetas.turing.ac.uk/sites/default/files/2024-07/cetas_briefing_paper_-_evaluating_malicious_generative_ai_capabilities.pdf

Ji, Y., Zhang, A., Chen, Z., Wei, M., Yu, Z., Zhang, X., & Han, L. (2024). Lightweight Road Damage Detection Algorithm based on the Improved YOLO Model. *2024 5th International Conference on Artificial Intelligence and Electromechanical Automation, AIEA 2024*, 832–835. https://doi.org/10.1109/AIEA62095.2024.10692408

Jiang, C., Zhou, Q., Lei, J., & Wang, X. (2022). A Two-Stage Structural Damage Detection Method Based on 1D-CNN and SVM. *Applied Sciences (Switzerland)*, *12*(20). https://doi.org/10.3390/app122010394

Jung, J., Kim, D. J., Vadivel, S. K. P., & Yun, S. H. (2019). Long-Term Deflection Monitoring for Bridges Using X and C-Band Time-Series SAR Interferometry. *Remote Sensing 2019, Vol. 11, Page 1258*, *11*(11), 1258. https://doi.org/10.3390/RS11111258

Kaur, N., Lee, C. C., Mostafavi, A., & Mahdavi-Amiri, A. (2023). Large-scale building damage assessment using a novel hierarchical transformer architecture on satellite images. *Computer-Aided Civil and Infrastructure Engineering*, *38*(15), 2072–2091. https://doi.org/10.1111/MICE.12981;PAGE:STRING:ARTICLE/CHAPTER

Khan, M. W., Obaidat, M. S., Mahmood, K., Batool, D., Badar, H. M. S., Aamir, M., & Gao, W. (2024). Real-Time Road Damage Detection and Infrastructure Evaluation Leveraging Unmanned Aerial Vehicles and Tiny Machine Learning. *IEEE Internet of Things Journal*, *11*(12), 21347–21358. https://doi.org/10.1109/JIOT.2024.3385994

Ko, B. C., Ham, S. J., & Nam, J. Y. (2011). Modeling and Formalization of Fuzzy Finite Automata for Detection of Irregular Fire Flames. *IEEE Transactions on Circuits and Systems for Video Technology*, *21*(12), 1903–1912. https://doi.org/10.1109/TCSVT.2011.2157190

Kopiika, N., Karavias, A., Krassakis, P., Ye, Z., Ninic, J., Shakhovska, N., Argyroudis, S., & Mitoulis, S.-A. (2025). Rapid post-disaster infrastructure damage characterisation using remote sensing and deep learning technologies: A tiered approach. *Automation in Construction*, *170*, 105955. https://doi.org/10.1016/J.AUTCON.2024.105955

Li, L., Sun, L., Ning, G., & Tan, S. (2014). Automatic Pavement Crack Recognition Based on BP Neural Network. *Promet - Traffic&Transportation*, *26*(1), 11–22. https://doi.org/10.7307/ptt.v26i1.1477

Li, Y. ;, Yin, C. ;, Lei, Y. ;, Zhang, J. ;, Yan, Y., Stefenon, F., Li, Y., Yin, C., Lei, Y., Zhang, J., & Yan, Y. (2024). RDD-YOLO: Road Damage Detection Algorithm Based on Improved You Only Look Once Version 8. *Applied Sciences 2024, Vol. 14, Page 3360*, *14*(8), 3360. https://doi.org/10.3390/APP14083360

Liu, C., Sui, H., Wang, J., Ni, Z., & Ge, L. (2022). Real-Time Ground-Level Building Damage Detection Based on Lightweight and Accurate YOLOv5 Using Terrestrial Images. *Remote Sensing*, *14*(12). https://doi.org/10.3390/rs14122763

Liu, M., Wang, Z., & Wang, P. (2018). Extension of D-TomoSAR for multi-dimensional reconstruction based on polynomial phase signal. *IET Radar, Sonar & Navigation*, *12*(4), 449–457. https://doi.org/10.1049/IET-RSN.2017.0450

Liu, W., Maruyama, Y., & Yamazaki, F. (2021). DAMAGE ASSESSMENT OF BRIDGES DUE TO THE 2020 JULY FLOOD IN JAPAN USING ALOS-2 INTENSITY IMAGESes. *International Geoscience and Remote Sensing Symposium (IGARSS)*, 3809–3812. https://doi.org/10.1109/IGARSS47720.2021.9554001

Lopez-Sanchez, A. ;, Multi-Annual, J. M., Villarroya-Carpio, A., & Lopez-Sanchez, J. M. (2023). Multi-Annual Evaluation of Time Series of Sentinel-1 Interferometric Coherence as a Tool for Crop Monitoring. *Sensors 2023, Vol. 23, Page 1833*, *23*(4), 1833. https://doi.org/10.3390/S23041833

Lu, C. H., Ni, C. F., Chang, C. P., Yen, J. Y., & Chuang, R. Y. (2018). Coherence difference analysis of sentinel-1 SAR interferogram to identify earthquake-induced disasters in urban areas. *Remote Sensing*, *10*(8). https://doi.org/10.3390/RS10081318

Ma, K., Hoai, M., & Samaras, D. (2017). Large-scale Continual Road Inspection: Visual Infrastructure Assessment in the Wild. *BMVC*.

Maeda, H., Sekimoto, Y., Seto, T., Kashiyama, T., & Omata, H. (2018). Road Damage Detection and Classification Using Deep Neural Networks with Smartphone Images. *Computer-Aided Civil and Infrastructure Engineering*, *33*(12), 1127–1141. https://doi.org/https://doi.org/10.1111/mice.12387

Majidifard, H., Jin, P., Adu-Gyamfi, Y., & Buttlar, W. G. (2020). Pavement Image Datasets: A New Benchmark Dataset to Classify and Densify Pavement Distresses. *Transportation Research Record*, *2674*(2), 328–339. https://doi.org/10.1177/0361198120907283

Mari, S., Valentini, G., Serva, S., Scopa, T., Cardone, M., Fasano, L., & De Luca, G. F. (2018). COSMO-SkyMed Second Generation System Access Portfolio. *IEEE Geoscience and Remote Sensing Magazine*, *6*(1), 35–43. https://doi.org/10.1109/MGRS.2017.2779461

Markert, K. N., Chishtie, F., Anderson, E. R., Saah, D., & Griffin, R. E. (2018). On the merging of optical and SAR satellite imagery for surface water mapping applications. *Results in Physics*, *9*, 275–277. https://doi.org/10.1016/J.RINP.2018.02.054

Markogiannaki, O., Xu, H., Chen, F., Mitoulis, S. A., & Parcharidis, I. (2022). Monitoring of a landmark bridge using SAR interferometry coupled with engineering data and forensics. *International Journal of Remote Sensing*, *43*(1), 95–119. https://doi.org/10.1080/01431161.2021.2003468

Mei, Q., & Gül, M. (2020). A cost effective solution for pavement crack inspection using cameras and deep neural networks. *Construction and Building Materials*, *256*, 119397. https://doi.org/10.1016/J.CONBUILDMAT.2020.119397

Motohka, T., Kankaku, Y., & Suzuki, S. (2017). Advanced Land Observing Satellite-2 (ALOS-2) and its follow-on L-band SAR mission. *2017 IEEE Radar Conference, RadarConf 2017*, 0953–0956. https://doi.org/10.1109/RADAR.2017.7944341

Mouzannar, H., Rizk, Y., & Awad, M. (2018, May 20). Damage Identification in Social Media Posts using Multimodal Deep Learning. *The 15th International Conference on Information*

Systems for Crisis Response and Management (ISCRAM). https://idl.iscram.org/files/husseinmouzannar/2018/2129_HusseinMouzannar_etal2018.pdf

Muhammad, K., Ahmad, J., & Baik, S. (2017). Early Fire Detection using Convolutional Neural Networks during Surveillance for Effective Disaster Management. *Neurocomputing*. https://doi.org/10.1016/j.neucom.2017.04.083

Mundt, M., Majumder, S., Murali, S., Panetsos, P., & Ramesh, V. (2019). Meta-learning Convolutional Neural Architectures for Multi-target Concrete Defect Classification with the COncrete DEfect BRidge IMage Dataset. *Proceedings of the IEEE Computer Society Conference on Computer Vision and Pattern Recognition*, *2019-June*, 11188–11197. https://doi.org/10.1109/CVPR.2019.01145

Nemni, E., Bullock, J., Belabbes, S., & Bromley, L. (2020). Fully Convolutional Neural Network for Rapid Flood Segmentation in Synthetic Aperture Radar Imagery. *Remote Sensing 2020, Vol. 12, Page 2532*, *12*(16), 2532. https://doi.org/10.3390/RS12162532

Nettis, A., Massimi, V., Nutricato, R., Nitti, D. O., Samarelli, S., & Uva, G. (2023). Satellite-based interferometry for monitoring structural deformations of bridge portfolios. *Automation in Construction*, *147*. https://doi.org/10.1016/j.autcon.2022.104707

Ni, Y., Mao, J., Fu, Y., Wang, H., Zong, H., & Luo, K. (2023). Damage Detection and Localization of Bridge Deck Pavement Based on Deep Learning. *Sensors 2023, Vol. 23, Page 5138*, *23*(11), 5138. https://doi.org/10.3390/S23115138

Niloy, F. F., Arif, Nayem, A. B. S., Sarker, A., Paul, O., Amin, M. A., Ali, A. A., Zaber, M. I., & Rahman, A. M. (2021). *A Novel Disaster Image Dataset and Characteristics Analysis using Attention Model*. https://doi.org/10.1109/ICPR48806.2021.9412504

Oliveira, H., & Correia, P. L. (2014). CrackIT — An image processing toolbox for crack detection and characterization. *2014 IEEE International Conference on Image Processing (ICIP)*, 798–802. https://doi.org/10.1109/ICIP.2014.7025160

Ouma, Y. O., & Hahn, M. (2017). Pothole detection on asphalt pavements from 2D-colour pothole images using fuzzy c-means clustering and morphological reconstruction. *Automation in Construction*, *83*, 196–211. https://doi.org/10.1016/J.AUTCON.2017.08.017

Pan, B., Tian, L., & Song, X. (2016). Real-time, non-contact and targetless measurement of vertical deflection of bridges using off-axis digital image correlation. *NDT & E International*, *79*, 73–80. https://doi.org/10.1016/J.NDTEINT.2015.12.006

Paramasivam, M. E., Perumal, S., & Pathmanaban, H. (2024). Revolutionizing Road Safety: AI-Powered Road Defect Detection. *2024 3rd International Conference on Power Electronics and IoT Applications in Renewable Energy and Its Control, PARC 2024*, 147–152. https://doi.org/10.1109/PARC59193.2024.10486759

Potnis, A. V., Shinde, R. C., Durbha, S. S., & Kurte, K. R. (2019). Multi-class segmentation of urban floods from multispectral imagery using deep learning. *International Geoscience and Remote Sensing Symposium (IGARSS)*, *2019-July*, 9741–9744. https://doi.org/10.1109/IGARSS.2019.8900250

Preiss, M., Gray, D. A., & Stacy, N. J. S. (2006). Detecting scene changes using synthetic aperture radar interferometry. *IEEE Transactions on Geoscience and Remote Sensing*, *44*(8), 2041–2054. https://doi.org/10.1109/TGRS.2006.872910

Radanliev, P. (2025). AI Ethics: Integrating Transparency, Fairness, and Privacy in AI Development. *Applied Artificial Intelligence*, *39*(1), 2463722. https://doi.org/10.1080/08839514.2025.2463722

Rahman, M. R., & Thakur, P. K. (2018). Detecting, mapping and analysing of flood water propagation using synthetic aperture radar (SAR) satellite data and GIS: A case study from the Kendrapara District of Orissa State of India. *The Egyptian Journal of Remote Sensing and Space Science*, *21*, S37–S41. https://doi.org/10.1016/J.EJRS.2017.10.002

Rebally, A., Valeo, C., He, J., & Saidi, S. (2021). Flood Impact Assessments on Transportation Networks: A Review of Methods and Associated Temporal and Spatial Scales. *Frontiers in Sustainable Cities*, *3*, 732181. https://doi.org/10.3389/FRSC.2021.732181/BIBTEX

Redmon, J., Divvala, S., Girshick, R., & Farhadi, A. (2016). You Only Look Once: Unified, Real-Time Object Detection. *Proceedings of the IEEE Conference on Computer Vision and Pattern Recognition*, 779–788. http://pjreddie.com/yolo/

Rizk, Y., Samer, H., Awad, M., & Castillo, C. (2019). A computationally efficient multi-modal classification approach of disaster-related Twitter images. *Association for Computing Machinery*, *F147772*, 2050–2059. https://doi.org/10.1145/3297280.3297481

Roy, D. P., Wulder, M. A., Loveland, T. R., C.E., W., Allen, R. G., Anderson, M. C., Helder, D., Irons, J. R., Johnson, D. M., Kennedy, R., Scambos, T. A., Schaaf, C. B., Schott, J. R., Sheng, Y., Vermote, E. F., Belward, A. S., Bindschadler, R., Cohen, W. B., Gao, F., … Zhu, Z. (2014). Landsat-8: Science and product vision for terrestrial global change research. *Remote Sensing of Environment*, *145*, 154–172. https://doi.org/10.1016/J.RSE.2014.02.001

Santaniello, P., & Russo, P. (2023). Bridge Damage Identification Using Deep Neural Networks on Time–Frequency Signals Representation. *Sensors*, *23*(13). https://doi.org/10.3390/s23136152

Sarwar, M. Z., & Cantero, D. (2024). Probabilistic autoencoder-based bridge damage assessment using train-induced responses. *Mechanical Systems and Signal Processing*, *208*, 111046. https://doi.org/10.1016/J.YMSSP.2023.111046

Schlögl, M., Widhalm, B., & Avian, M. (2021). Comprehensive time-series analysis of bridge deformation using differential satellite radar interferometry based on Sentinel-1. *ISPRS Journal of Photogrammetry and Remote Sensing*, *172*, 132–146. https://doi.org/10.1016/J.ISPRSJPRS.2020.12.001

Schultz, L. A. (2021). Synthetic Aperture Radar (SAR) RGB Quick Guide. *HydroSAR Training Event with ICIMOD-Supporting A. 33 ROSES/SERVIR AST*.

Shakhovska, N., Yakovyna, V., Mysak, M., Mitoulis, S. A., Argyroudis, S., & Syerov, Y. (2024). Real-Time Monitoring of Road Networks for Pavement Damage Detection Based on Preprocessing and Neural Networks. *Big Data and Cognitive Computing 2024, Vol. 8, Page 136*, *8*(10), 136. https://doi.org/10.3390/BDCC8100136

Sharma, R. C., Tateishi, R., Hara, K., Nguyen, H. T., Gharechelou, S., & Nguyen, L. V. (2017). Earthquake Damage Visualization (EDV) Technique for the Rapid Detection of Earthquake-Induced Damages Using SAR Data. *Sensors 2017, Vol. 17, Page 235*, *17*(2), 235. https://doi.org/10.3390/S17020235

Shi, Y., Cui, L., Qi, Z., Meng, F., & Chen, Z. (2016). Automatic Road Crack Detection Using Random Structured Forests. *IEEE Transactions on Intelligent Transportation Systems*, *17*(12), 3434–3445. https://doi.org/10.1109/TITS.2016.2552248

Sousa, H., Cavadas, F., Henriques, A., Bento, J., & Figueiras, J. (2013). Bridge deflection evaluation using strain and rotation measurements. *Smart Structures and Systems*, *11*(4), 365–386. https://doi.org/10.12989/sss.2013.11.4.365

Space Agency, E. (2012a). *Sentinel-1 eSA's Radar Observatory Mission for GMeS Operational Services*. www.esa.int

Space Agency, E. (2012b). *Sentinel-2 eSA's Optical High-Resolution Mission for GMeS Operational Services*. www.esa.int

Stricker, R., Aganian, D., Sesselmann, M., Seichter, D., Engelhardt, M., Spielhofer, R., Hahn, M., Hautz, A., Debes, K., & Gross, H.-M. (2021). *Road Surface Segmentation - Pixel-Perfect Distress and Object Detection for Road Assessment*. https://doi.org/10.1109/CASE49439.2021.9551591

Stricker, R., Eisenbach, M., Sesselmann, M., Debes, K., & Gross, H. M. (2019). Improving Visual Road Condition Assessment by Extensive Experiments on the Extended GAPs Dataset. *Proceedings of the International Joint Conference on Neural Networks*, *2019-July*. https://doi.org/10.1109/IJCNN.2019.8852257

Sun, X., Zimmer, A., Mukherjee, S., Kottayil, N. K., Ghuman, P., & Cheng, I. (2020). DeepInSAR—A Deep Learning Framework for SAR Interferometric Phase Restoration and Coherence Estimation. *Remote Sensing 2020, Vol. 12, Page 2340*, *12*(14), 2340. https://doi.org/10.3390/RS12142340

Tazarv, M., Won, K., Jang, Y., Hart, K., & Greeneway, E. (2022). Post-earthquake serviceability assessment of standard RC bridge columns using computer vision and seismic analyses. *Engineering Structures*, *272*, 115002. https://doi.org/10.1016/J.ENGSTRUCT.2022.115002

Teopilus, C. D., & Amrozi, M. R. F. (2023). The Evaluation of Pavement Condition Assessment Methods for Road Assets in Coastal Areas. *INERSIA Lnformasi Dan Ekspose Hasil Riset Teknik Sipil Dan Arsitektur*, *19*(2), 183–193. https://doi.org/10.21831/inersia.v19i2.61089

Tian, L., Zhao, J., Pan, B., Wang, Z., Kohut, P., Sabato, A., Martowicz, A., & Holak, K. (2021). Full-Field Bridge Deflection Monitoring with Off-Axis Digital Image Correlation. *Sensors 2021, Vol. 21, Page 5058*, *21*(15), 5058. https://doi.org/10.3390/S21155058

Tonelli, D., Caspani, V., Valentini, A., Rocca, A., Torboli, R., Vitti, A., Perissin, D., & Zonta, D. (2023). Interpretation of Bridge Health Monitoring Data from Satellite InSAR Technology. *Remote Sensing*, *15*, 5242. https://doi.org/10.3390/rs15215242

United Nations. (2015). *THE 17 GOALS | Sustainable Development*. Https://Sdgs.Un.Org/Goals. https://sdgs.un.org/goals

Verstockt, S., Beji, T., De Potter, P., Van Hoecke, S., Sette, B., Merci, B., & Van de Walle, R. (2013). Video driven fire spread forecasting (f) using multi-modal LWIR and visual flame and smoke data. *Pattern Recognition Letters*, *34*(1), 62–69. https://doi.org/https://doi.org/10.1016/j.patrec.2012.07.018

V.V., D., O.P., G., & I.O., K. (2024). Application of convolutional neural networks to detect damaged buildings. *Sistemi Ta Tehnologìï*, *3*(152), 107–114. https://doi.org/10.34185/1562-9945-3-152-2024-11

Wang, C., Liu, Y., Zhang, X., Li, X., Paramygin, V., Sheng, P., Zhao, X., & Xu, S. (2024). Scalable and rapid building damage detection after hurricane Ian using causal Bayesian networks and InSAR imagery. *International Journal of Disaster Risk Reduction*, *104*, 104371. https://doi.org/10.1016/J.IJDRR.2024.104371

Wang, Y., Chew, A. W. Z., & Zhang, L. (2022). Building damage detection from satellite images after natural disasters on extremely imbalanced datasets. *Automation in Construction*, *140*, 104328. https://doi.org/10.1016/J.AUTCON.2022.104328

Waseem Khan, M., Obaidat, M. S., Mahmood, K., Sadoun, B., Sanaullah Badar, H. M., & Gao, W. (2025). Real-Time Road Damage Detection Using an Optimized YOLOv9s-Fusion in IoT Infrastructure. *IEEE Internet of Things Journal*. https://doi.org/10.1109/JIOT.2025.3537640

Weber, E., & Kané, H. (2020). *Building Disaster Damage Assessment in Satellite Imagery with Multi-Temporal Fusion*. https://arxiv.org/abs/2004.05525v1

Weber, E., Papadopoulos, D. P., Lapedriza, A., Ofli, F., Imran, M., & Torralba, A. (2023). Incidents1M: A Large-Scale Dataset of Images With Natural Disasters, Damage, and Incidents. *IEEE Transactions on Pattern Analysis and Machine Intelligence*, *45*, 4768–4781. https://doi.org/10.1109/TPAMI.2022.3191996

Weng, X., Huang, Y., & Wang, W. (2019). Segment-based pavement crack quantification. *Automation in Construction*, *105*, 102819. https://doi.org/10.1016/J.AUTCON.2019.04.014

Werninghaus, R., & Buckreuss, S. (2010). The TerraSAR-X mission and system design. *IEEE Transactions on Geoscience and Remote Sensing*, *48*(2), 606–614. https://doi.org/10.1109/TGRS.2009.2031062

*What is Data Governance? | IBM*. (2025). IBM. https://www.ibm.com/think/topics/data-governance

*What is Explainable AI (XAI)? | IBM*. (2025). IBM. https://www.ibm.com/think/topics/explainable-ai

Winsvold, S. H., Kääb, A., Nuth, C., Andreassen, L. M., Van Pelt, W. J. J., & Schellenberger, T. (2018). Using SAR satellite data time series for regional glacier mapping. *Cryosphere*, *12*(3), 867–890. https://doi.org/10.5194/TC-12-867-2018

Wu, C., Zhang, F., Xia, J., Xu, Y., Li, G., Xie, J., Du, Z., & Liu, R. (2021). Building Damage Detection Using U-Net with Attention Mechanism from Pre- and Post-Disaster Remote Sensing Datasets. *Remote Sensing 2021, Vol. 13, Page 905*, *13*(5), 905. https://doi.org/10.3390/RS13050905

Wu, R., Liu, G., Zhang, R., Wang, X., Li, Y., Zhang, B., Cai, J., & Xiang, W. (2020). A Deep Learning Method for Mapping Glacial Lakes from the Combined Use of Synthetic-Aperture Radar and Optical Satellite Images. *Remote Sensing 2020, Vol. 12, Page 4020*, *12*(24), 4020. https://doi.org/10.3390/RS12244020

Wu, Y., Li, J., Wu, Y., & Li, J. (2021). Deflection Measurement for Bridges Based on Secant Inclination. *Open Journal of Civil Engineering*, *11*(4), 427–433. https://doi.org/10.4236/OJCE.2021.114025

Xu, H., Su, X., Wang, Y., Cai, H., Cui, K., & Chen, X. (2019). Automatic Bridge Crack Detection Using a Convolutional Neural Network. *Applied Sciences 2019, Vol. 9, Page 2867*, *9*(14), 2867. https://doi.org/10.3390/APP9142867

Xu, J. Z., Lu, W., Li, Z., Khaitan, P., & Zaytseva, V. (2019). Building Damage Detection in Satellite Imagery Using Convolutional Neural Networks. *ArXiv.Org*.

Yang, F., Zhang, L., Yu, S., Prokhorov, D., Mei, X., & Ling, H. (2020). Feature Pyramid and Hierarchical Boosting Network for Pavement Crack Detection. *IEEE Transactions on Intelligent Transportation Systems*, *21*(4), 1525–1535. https://doi.org/10.1109/TITS.2019.2910595

Yang, Y., Xie, C., Tian, B., Guo, Y., Zhu, Y., Yang, Y., Fang, H., Bian, S., & Zhang, M. (2024). Large-scale building damage assessment based on recurrent neural networks using SAR coherence time series: A case study of 2023 Turkey–Syria earthquake. *Earthquake Spectra*. https://doi.org/10.1177/87552930241262761/SUPPL_FILE/SJ-DOCX-1-EQS-10.1177_87552930241262761.DOCX

Yun, S. H., Hudnut, K., Owen, S., Webb, F., Simons, M., Sacco, P., Gurrola, E., Manipon, G., Liang, C., Fielding, E., Milillo, P., Hua, H., & Coletta, A. (2015). Rapid Damage Mapping for the 2015 Mw 7.8 Gorkha Earthquake Using Synthetic Aperture Radar Data from COSMO–SkyMed and ALOS-2 Satellites. *Seismological Research Letters*, *86*(6), 1549–1556. https://doi.org/10.1785/0220150152

Yunmei, J., Huifeng, W., Haoyi, C., Bei, Y., Zhihui, H., Shangzhen, S., Limin, W., & He, H. (2023). Multi-point detection method of dynamic deflection of super long-span bridge based on chain laser model. *Measurement*, *209*, 112535. https://doi.org/10.1016/J.MEASUREMENT.2023.112535

Zanevych, Y., Yovbak, V., Basystiuk, O., Shakhovska, N., Fedushko, S., & Argyroudis, S. (2024). Evaluation of Pothole Detection Performance Using Deep Learning Models Under Low-Light Conditions. *Sustainability 2024, Vol. 16, Page 10964*, *16*(24), 10964. https://doi.org/10.3390/SU162410964

Zeng, J., & Zhong, H. (2024). YOLOv8-PD: an improved road damage detection algorithm based on YOLOv8n model. *Scientific Reports 2024 14:1*, *14*(1), 1–14. https://doi.org/10.1038/s41598-024-62933-z

Zhang, L., Yang, F., Zhang, Y. D., & Zhu, Y. J. (2016). Road crack detection using deep convolutional neural network. *2016 IEEE International Conference on Image Processing (ICIP)*, 3708–3712. https://doi.org/10.1109/ICIP.2016.7533052

Zhang, W., Sun, L. M., & Sun, S. W. (2017). Bridge-Deflection Estimation through Inclinometer Data Considering Structural Damages. *Journal of Bridge Engineering*, *22*(2), 04016117. https://doi.org/10.1061/(ASCE)BE.1943-5592.0000979/ASSET/A0FE8353-4526-4AD6-9131-FBD17B380626/ASSETS/IMAGES/LARGE/FIGURE16.JPG

Zhang, Y., & Prinet, V. (2004). InSAR coherence estimation. *International Geoscience and Remote Sensing Symposium (IGARSS)*, *5*, 3353–3355. https://doi.org/10.1109/IGARSS.2004.1370422

Zhang, Y., Wang, Z., Luo, Y., Yu, X., & Huang, Z. (2023). Learning Efficient Unsupervised Satellite Image-based Building Damage Detection. *2023 IEEE International Conference on Data Mining (ICDM)*, 1547–1552. https://api.semanticscholar.org/CorpusID:265608725

Zhao, X., & Morikawa, S. (2024). Rapid assessment of large-scale urban destruction in conflict zones using hypergraph-based visual-structural machine learning. *Journal of Engineering Research*. https://doi.org/10.1016/J.JER.2024.08.006

Zou, Q., Cao, Y., Li, Q., Mao, Q., & Wang, S. (2012). CrackTree: Automatic crack detection from pavement images. *Pattern Recognition Letters*, *33*(3), 227–238. https://doi.org/10.1016/J.PATREC.2011.11.004